\documentclass[aip,reprint,superscriptaddress]{revtex4-2}

\usepackage{graphicx}
\usepackage{amsmath,amsfonts,amssymb}
\usepackage{bm}

\usepackage{xcolor}
\usepackage[colorlinks=true,linkcolor=blue,citecolor=blue,urlcolor=blue]{hyperref}

\usepackage[caption=false]{subfig}

\usepackage{float}
\usepackage{url}
\usepackage{cleveref}

\begin{document}
%
\title{Resonant excitation of terahertz surface magnetoplasmons by two p-polarized beating lasers interacting on a graphene-n-InSb surface}

	\author{Rohit Kumar Srivastav}%
	\affiliation{\mbox{Institute For Plasma Research, Bhat,
		Gandhinagar 382 428, Gujarat, India}}%
	\affiliation{\mbox{Department of Physics, JSS University, Noida-62, Gautam Buddha Nagar, Uttar Pradesh, 201301, India}}%

	\author{Mrityunjay Kundu}%
	\email{mkundu@ipr.res.in}
	\affiliation{\mbox{Institute For Plasma Research, Bhat, Gandhinagar 382 428,
		Gujarat, India}}%
	\affiliation{\mbox{Homi Bhabha National Institute, Training School Complex,
		Anushaktinagar, Mumbai 400094, India}}%
\date{\today}
\begin{abstract}
A mechanism of resonant excitation of
surface magnetoplasmons (SMPs) is proposed in the terahertz (THz) frequency range by {\emph{beating of two
p-polarized lasers}}, obliquely incident at an angle $\theta$ on a graphene
sheet deposited over a rippled surface of a magnetized n-type semiconductor. The 
resulting laser-beat-envelope
induces a nonlinear velocity to free electrons, which
couples with the modulated charge carrier density and
generates a nonlinear current. This time-varying oscillating nonlinear
current acts as the source of THz SMPs
wave generation, as opposed to THz generation by a different process with {\emph{a single laser}} in the earlier work [Phys. Rev. E 113, 015208 (2026)] where light dispersion characteristics as well as the required phase-matching conditions are markedly different.
The resulting THz SMPs field amplitude
is shown to be controlled in the frequency range of $2-5$~THz by varying the graphene's Fermi energy
($\textrm{E}_\textrm{F}=20-130$ meV), laser incident angle ($\theta = 0-90^{o}$), the semiconductor's temperature ($T = 320 - 380$~K) and external
magnetic field ($\textrm{B}_{0} \approx 0 - 0.09 $~T).
The amplitude of THz SMPs field now reaches on the order of $10^{-1}$
w.r.t. the incident field amplitude, and it is almost $10^1 - 10^2$ fold higher compared to previous works.
Thus, the proposed mechanism may open new avenues
for the development of actively tunable plasmonic device,
with potential applications in future THz technologies and 6G wireless communication systems.
\end{abstract}
%
%
%
\maketitle
\vspace{-0.25cm}
\section{Introduction}
\vspace{-0.25cm}
Plasmonics has emerged as a vibrant and rapidly advancing field in recent 
times due to its unique optical and physical properties. These include highly 
enhanced light–matter interactions, strong resonance phenomena, sub-wavelength 
electromagnetic field confinement, and exceptional sensitivity to surface 
geometries and perturbations\cite{liu2024electronic}. Such characteristics have 
enabled transformative advancements across multiple disciplines, including 
spectroscopy\cite{nenninger2002data}, biomedical imaging and 
diagnostics\cite{masson2017surface}, high-speed communication\cite{10106756}, 
and precision sensing technologies\cite{dong2024enhanced}.
At the core of many plasmonic phenomena remains surface plasmon polaritons 
(SPPs), which are evanescent electromagnetic waves generated at the interface 
between materials with contrasting dielectric properties typically 
metal-dielectric or semiconductor-dielectric 
boundaries\cite{pitarke2006theory,zhang2012surface,han2012radiation}. These waves originate from the interaction between incident light and the collective electron oscillations at the interface. A variety of excitation 
mechanisms have been proposed and studied for the generation of SPPs, including 
harmonic generation\cite{de2016harmonics,sedaghat2024giant}, filamentation 
instabilities\cite{kumar2007filamentation}, optical 
rectification\cite{srivastav2024resonant}, Cerenkov-type 
interactions\cite{liu2024excitation}, and the beating of dual laser 
frequencies\cite{chamoli2024surface}. Furthermore, in the presence of magnetic 
field surface plasmons (SPs) becomes surface magnetoplasmons (SMPs). It has been 
actively explored in the terahertz (THz) frequency regime, offering promising 
avenues for compact THz photonic and plasmonic device architectures.

 Numerous mechanisms involve THz wave generation, including laser–plasma, electron beam–plasma, laser–material, and electron beam–material interactions\cite{molavi2025role,mishra2025twisted,yun2025laser,chamoli2025optical,SrivastavNLMSH23,zhang2024plane}. The generation of THz wave through laser–material (metal, semiconductor or graphene) interactions is of significant importance due to its highly tunable nature\cite{liu2015directional,srivastav2023excitation}. By manipulating both laser parameters and the intrinsic properties of the target material, one can control the spectral, temporal, and spatial characteristics of the emitted wave. This controllability renders laser-driven sources highly versatile for diverse applications, including optical communication, biomedical imaging, sensing, and detection systems\cite{korani2024tunable,vatoor2025tunable,ukirade2025review,hasan2024terahertz}.

Kumar \emph{et.al.}\cite{kumar2022combined} examined the impact of transverse electric and magnetic field over the THz wave generation in the collisional plasma using the beating of two amplitude modulated lasers.
Ghayemmoniri \emph{et.al.}\cite{ghayemmoniri2023terahertz} studied the THz wave generation in the presence of trapped magnetic field over carbon nanotube using a nonlinear mixing (NLM) of chirped lasers. 
Javan and Erid\cite{sepehri2017theoretical} investigated THz wave generation by NLM of two Gaussian lasers over spherical graphite nanoparticle and THz wave amplitude of the order of $10^{-5}$. Kolur and Mahdi Esmaeilzadeha\cite{rahmanpour2025terahertz} examined the effect of wiggler magnetic field over THz wave generation by NLM of two super Gaussian lasers in a collisional and rippled density plasma. Kumar\emph{ et.al.}\cite{10456555} reported THz wave generation by NLM of two lasers one is Cosh-Gaussian laser another is dark hollow Gaussian laser in rippled density under dense plasma and they observe that the larger THz wave amplitude as compare to the NLM of two Gaussian lasers.
Singh  \emph{et.al}  \cite{singh2024laser} investigated THz wave generation by
NLM of two flat Gaussian pulses on the n-InSb semiconductor. Chamoli \emph{et.al} \cite{chamoli2024surface} studied the THz
SMPs wave generation on a metal surface in the presence of an
an externally applied static magnetic field and used a magnetic field range from 10 T
to 30 T. Srivastav and Panwar\cite{srivastav2022excitation}
examined analytically THz SMPs wave generation over n-InSb
semiconductor in the presence of an externally applied static magnetic field and
reported normalized amplitude of THz SMPs of the order of $\approx10^{-3}$ with 
an externally applied static magnetic field $\approx3750$ Gauss. Recently, Srivastav and 
Kundu\cite{srivastav2025terahertz} explored the generation of THz SMPs wave 
through {\emph{linear mode conversion}} (LMC) of THz waves at the 
graphene–n-InSb 
semiconductor 
interface in the presence of an externally applied static magnetic field. Their earlier 
LMC approach~\cite{srivastav2025terahertz} 
utilized a single laser. The present study, however, employs two 
lasers to generate a beat envelope enabling THz 
SMPs wave by {\emph{a different physical mechanism}} described below.

In the present paper, we investigate the generation of THz SMPs wave through the 
NLM of two obliquely incident p-polarized lasers at an angle 
$\theta$ on a rippled graphene-n-type semiconductor (n-InSb) surface, with an 
externally applied static magnetic field strength, $\vec{B} = B_0 \hat{y}$ applied to the n-InSb 
semiconductor. The incident lasers, with frequencies $\omega_{1}$ and 
$\omega_{2}$, impart linear oscillatory velocity to the free electrons of the 
graphene-n-InSb semiconductor and also exert a nonlinear ponderomotive force on 
the free electrons at the frequency $\omega = \omega_{1} - 
\omega_{2}$. Under the action of the ponderomotive force, electrons acquire a nonlinear oscillatory velocity $\vec{V}_{\omega}$ that combines with the modulated electron density $n_q$ to produce the oscillating nonlinear current density $\vec{J}_{\omega}$. This 
$\vec{J}_{\omega}$ term resonantly drives the THz SMPs wave at frequency $\omega$ with a 
propagation constant $k_z = k_{1z} - k_{2z} + q$, where $q$ provides the 
additional momentum for the generation of the THz SMPs wave.

The paper is organized as follows. In Sec.~\ref{sec2}, the analytical expression for the nonlinear current density is derived. Section~\ref{sec3} determines the field amplitude of the THz SMPs wave. The results are discussed in Sec.~\ref{sec4}, and the conclusions are given in Sec.~\ref{sec5}.

\vspace{-0.5cm}
\section{Oscillating nonlinear current density: origin of THz wave}\label{sec2}
\begin{figure}
		\centering
		\includegraphics[width=0.45\textwidth]{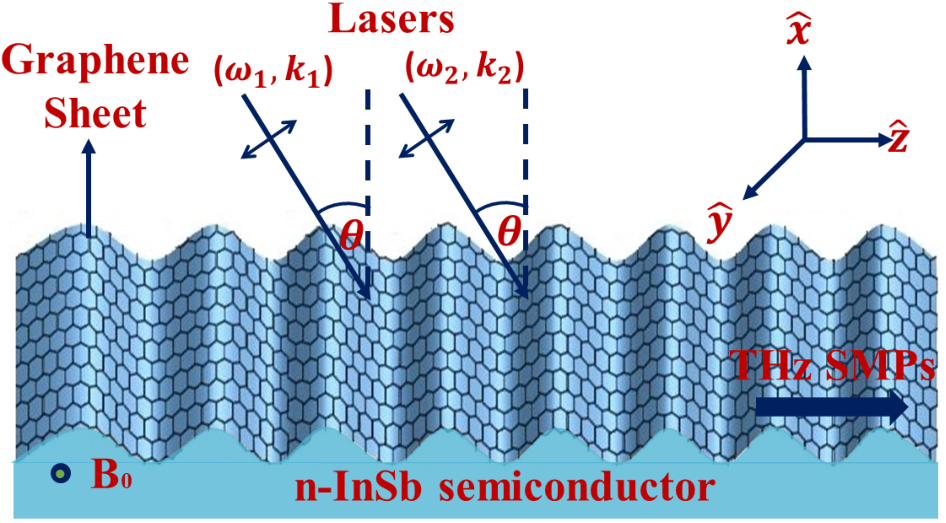}
		\caption{\label{Fig1} Schematic illustration of THz SMPs wave generation via NLM of two obliquely incident p-polarized laser beams on a graphene–n-InSb rippled surface.}
\vspace{-0.5cm}
	\end{figure}

\vspace{-0.25cm}
	Let us consider a two p-polarized laser beams with frequencies $\omega_{j}$ and wave numbers 
$k_{jx}$,
$k_{jz}$ along $\hat{x}$ and $\hat{z}$ respectively, where $j=1,2$, incident at
an angle $\theta$ on the rippled graphene-n-InSb surface as depicted in Figure.\ref{Fig1}. Here, the rippled surface provides perturbation in the electron density with $n=n_{q}+(n_{0}/2)$, $hq\le1$
here $n_{q} = (n_{0}\cos qz)/2$, with $h$ and $q$ representing the ripple amplitude and wave number, respectively~\cite{singh2007surface,bhasin2010resonant}.
	If the graphene's Fermi energy $\text{E}_\text{F} \gg \hbar 
	\omega_{j}$, then graphene
	conductivity $\sigma_{j,g} (= (i {\mathrm{e}}^2 \text{E}_\text{F}) / (\pi \hbar^2 
	(\omega_{j} + i \nu)))$ lies in THz frequency regime. Here $\nu$, $\hbar$ and $\mathrm{e}$ are the  average collision frequency of electrons, reduced Planck constant and charge of an electron, respectively. The graphene Fermi level can be modulated via an externally applied gate voltage~\cite{yarahmadi2015subwavelength,liu2011graphene}. An externally applied magnetic field $B_0$ is applied in the $\hat{y}$ direction over the n-InSb medium. Under an externally applied magnetic field, n-InSb's permittivity becomes a
	tensor~$\overset{=}{\epsilon}$\cite{brion1972theory};
with components
	$\epsilon_{x x} = \epsilon_{z z} = \epsilon_r -  (\epsilon_r\omega_p^2 (\omega_{j}
	+ i \nu)) /( \omega_{j} ((\omega_{j} + i \nu)^2 - \omega_{{ce}}^2))$, $\epsilon_{x
		z} = - \epsilon_{z x} = -  (i \epsilon_r\omega_p^2 \omega_{{ce}}) /
	(\omega_{j} ((\omega_{j} + i \nu)^2 - \omega_{{ce}}^2))$, $\epsilon_{y y} =
	\epsilon_r -  (\epsilon_r\omega_p^2 )/( \omega_{j} (\omega_{j} + i
	\nu))$. Here, $\omega_{ce}$ ($= e B_0 / m^{\ast}_e$) is the electron cyclotron frequency, $\omega_p = \sqrt{n_0 e^2 / m^{\ast}_e \epsilon_r
		\epsilon_0}$ is the electron plasma frequency, $m^{\ast}_e = 0.014 m_e $ denotes the
	effective free electron mass, and $\epsilon_r$ denotes the permittivity of n-InSb. The free electron
	charge density $n_0  = 5.76 \times 10^{20} T^{3 / 2} \exp(-{E_{g}}/2k_{B} T)\,\, (\textrm{in m}^{-3})$ is related to the band gap energy $E_{g} = 0.26 \text{eV}$, temperature $T (\text{in kelvin})$ of n-InSb  and the Boltzmann constant $k_B$ (in eV/Kelvin)~\cite{jing2022thermally, gao2023multifunctional}. The plasma is assumed to be under-dense, i.e., $\omega_{j}>\omega_{p}$.
    
The incident p-polarized lasers (see Figure.\ref{Fig1}) electric field corresponding to the amplitude
$E_j$ follows
\vspace{-0.2cm}
\begin{equation}
	\vec{E_{j}} = E_{j} (\hat{z} + \tan \theta \hat{x}) e^{- i (\omega_{j} t + k_{j x} x -
		k_{j z} z)}\label{eq1} 
\end{equation}

\noindent
where $E_{j} = E_{0,j} \cos \theta$, $k_{j x} = k_j \cos \theta$, 
$k_{j z} = k_j \sin
\theta$, $k_j = \omega_{j}/ c = \sqrt{k_{jx}^{2} + k_{jz}^{2}}$ is wave vector amplitude in free space, and $c$ represents the speed of light.
%
%
The incident electromagnetic wave is partially reflected and partially transmitted. The transmitted wave’s electric field reads
\vspace{-0.1cm}
\begin{equation}
	\vec{E}_{j,tr} = E_{j} T_{j,tr} (\hat{z} + \beta_{j} \hat{x}) e^{\alpha_{j} x} e^{- i (\omega_{j} t -
		k_{j z} z)}\label{eq2}
\end{equation}

\noindent 
where $\beta_{j} = (\epsilon_{j,x z} \alpha_{j}  + \epsilon_{j,x x} i k_{j z}) / 
(-
\epsilon_{j,x x} \alpha_{j}+ \epsilon_{j,x z} {i k_{j z}} )$ for $x \leq 0$,
$\alpha_{j}^2 = k^2_{j z} - (\omega_{j}^2 / c^2) \epsilon_{j,eff}$,
$\epsilon_{j,eff} = (\epsilon^2_{j,x x} + \epsilon_{j,x z}^2) /
\epsilon_{j,x x}$ and $T_{j,tr} = (2 \epsilon_0 \cos \theta) / ((1 +
(\epsilon_{j,eff} \beta_{j}) / \tan \theta) \epsilon_0 - (\sigma_{j,g} \cos
\theta) / c)$ is the transmission coefficient\cite{srivastav2025terahertz}. 
The effect of graphene enters through $\sigma_{j,g}$ via 
the transmission coefficient $T_{tr}$.

The interaction of the incident laser beam with the graphene–n-InSb semiconductor surface leads to the ionization of atoms and the production of free electrons. Using the equation of motion $m_{e}^{\ast}[\partial(\vec{V}_{j})/\partial{t}+\vec{V}_{j}\cdot(\vec{\nabla}\vec{V}_{j})]=-e[\vec{E}+\vec{V}_{j}\times\vec{B}]$\cite{chen1984introduction} along with Eq. \eqref{eq2}, the linear oscillatory velocity of the free electrons can be written as,
\vspace{-0.25cm}
\begin{equation}
	\vec{V}_{j} = \frac{e}{m_e^{\ast}} (\tilde{\textrm{v}}_{j,x}  \hat{x} +
	\tilde{\textrm{v}}_{j,z}  \hat{z}) {T_{j,tr} E_j}  e^{\alpha_{j} x} e^{- i \left( \omega_{j} 
		{t - k_{j z}}  z \right)}\label{eq3}
 		\vspace{-0.15cm}
\end{equation}
where, $\tilde{\textrm{v}}_{j,x} = (\omega_{ce} - i (\omega_{j} + i \nu) 
\beta_{j}) / ((\omega_{j} + i \nu)^2 - \omega_{ce}^2)$ and
$\tilde{\textrm{v}}_{j,z} = (-\omega_{ce} \beta_{j} - i (\omega_{j} + i \nu) ) /
((\omega_{j} + i \nu)^2 - \omega_{ce}^2)$.
Lasers exert a nonlinear ponderomotive force $ \vec{F}_p = ({1/2}){Re} [- m_e^{\ast} (\vec{V}_j \cdot
\vec{\nabla}) \vec{V}^{\ast}_j - e (\vec{V}_j \times 
\vec{B}^{\ast})]$\cite{liu2019high} on free electrons at their difference 
frequency $\omega = \omega_{1}-\omega_{2}$,
\vspace{-0.25cm}
\begin{equation}
	\vec{F}_p = {- \frac{e^2}{4 m_e^{\ast}}} \left(\tilde{F}^x_p
	{\hat{x}} + \tilde{F}^z_p {\hat{z}}\right)\tilde{E}\label{eq4} 
	\vspace{-0.25cm}
\end{equation}
where,  $\tilde{E}=T_{1,tr} E_1 T_{2,tr}^{\ast}
E^{\ast}_2 e^{(\alpha_1 + \alpha_2^{\ast} ) } e^{- i (\omega t - (k_{1z} - k_{2 z}) z)}$, $\tilde{F}^x_p=((\tilde{\textrm{v}}_{1,x} \alpha_2 -
\tilde{\textrm{v}}_{1,z} i k_{2 z}) (\tilde{\textrm{v}}_{2,x})^{\ast} +
((\tilde{\textrm{v}}_{2,x})^{\ast} \alpha_1 + (\tilde{\textrm{v}}_{2,z})^{\ast}
i k_{1 z}) \tilde{\textrm{v}}_{1,x} )  +(({(\tilde{\textrm{v}}_{2,z})^{\ast} 
	(\alpha_1 - \beta_1 i k_{1 z})}/{i\omega_1}) - ({\tilde{\textrm{v}}_{1,z} 
(\alpha^{\ast}_2 + i \beta_2 k_{2 z})}/{i \omega_2}))$ and 
$\tilde{F}^z_p=((\tilde{\textrm{v}}_{1,x} \alpha_2 - \tilde{\textrm{v}}_{1,z} i 
k_{2 z}) (\tilde{\textrm{v}}_{1,z})^{\ast} + ((\tilde{\textrm{v}}_{2,x})^{\ast} 
\alpha_1 + (\tilde{\textrm{v}}_{2,z})^{\ast} i k_{1 z}) 
\tilde{\textrm{v}}_{1,z})\\ + ( ( \tilde{v}_{1,x} (\alpha^{\ast}_2 + i \beta_2 
k_{2 z})/{i \omega_2}) - (({(\tilde{\textrm{v}}_{2,x})^{\ast} (\alpha_1 - 
\beta_1 i k_{1 z})})/{i \omega_1}))$.

Electrons acquire an oscillatory velocity $\vec{V}_{\omega}$ under the action of the ponderomotive force at $\omega = \omega_{1}-\omega_{2}$,
\vspace{-0.25cm}
	\begin{equation}
		\vec{\textrm{V}}_{\omega} = - \frac{e^2}{4 (m_e^{\ast})^2} \left(\tilde{\textrm{V}}^{x}_\omega \hat{x} + \tilde{\textrm{V}}^{z}_\omega \hat{z}\right)\tilde{E}\label{eq5}
		\vspace{-0.25cm}
	\end{equation}
%
%
where, $\tilde{\textrm{V}}^{x}_\omega = (- \omega_{{ce}} \tilde{F}^z_p
		+ i (\omega + i \nu) \tilde{F}^x_p)/( (\omega + i \nu)^2 -
		\omega_{{ce}}^2 ) $ and $\tilde{\textrm{V}}^{z}_\omega = (\omega_{{ce}} \tilde{F}^x_p +
		i (\omega + i \nu) \tilde{F}^z_p)/((\omega + i \nu)^2 -
		\omega_{{ce}}^2)$.
In the rippled regime, the nonlinear current density $\vec J^{nl}_{\omega} =
-n_{q} e \vec V_{\omega}$ develops at the frequency difference $\omega =
\omega_{1}-\omega_{2}$ and modified propagation wave number $k_z = k_{1z} -
k_{2z} + q$. This $\vec J^{nl}_{\omega}$ can be
expressed as,
\vspace{-0.25cm}
\begin{equation}
	\vec{J}^{n l}_{\omega} = \frac{n_0 e^3}{8
		(m_e^{\ast})^2} \left(\tilde{J}_{\omega}^x \hat{x} + \tilde{J}_{\omega}^z \hat{z}\right)\tilde{\tilde{E}}\label{eq6} 
		\vspace{-0.25cm}
\end{equation}
where, $\tilde{\tilde{E}} = T_{1,tr} E_1 T_{2,tr}^{\ast} E^{\ast}_2 e^{(\alpha_1 + \alpha_2^{\ast} ) } e^{- i (\omega t - k_{z} z)}, \tilde{J}^{nl}_{\omega,x} = ({- \omega_{{ce}} \tilde{F}^z_p + i (\omega + i \nu) \tilde{F}^x_p})/({ (\omega + i \nu)^2 -
	\omega_{{ce}}^2 })$ and $\tilde{J}^{nl}_{\omega,z} = ({\omega_{{ce}} \tilde{F}^x_p + i
	(\omega + i \nu) \tilde{F}^z_p}) /({ (\omega + i \nu)^2 -
	\omega_{{ce}}^2 })$. The THz SMPs wave is excited by the term $\vec{J}^{nl}_{\omega}$, acting as a source.

	In passing, one may note that, this present mechanism of THz wave 
generation {\emph{is completely different}} than the previously reported 
LMC process\cite{srivastav2025terahertz}. Also, the respective wave-dispersion 
relations (shown in the Appendix), phase-matching conditions, THz SMPs field
amplitude are {\emph{markedly different}}.

\vspace{-0.4cm}
\section{\!\!\!The magnitude of THz SMPs electric field }\label{sec3}
\vspace{-0.25cm}
The THz SMPs wave is driven by the $\vec{J}_{\omega}^{n l}$ at ($\omega, k_{z})$ within the rippled region.
 Consider a self-consistent THz SMPs field $\vec{E}_{SMPs}$, within the rippled regime of the graphene–n-InSb. Using Maxwell’s equations $\vec{\nabla} \times \vec{E}_{SMPs} = - (\partial
\vec{B}_{SMPs}/\partial t)$ and $\vec{\nabla} \times
\vec{B}_{SMPs} = \mu_0 \vec{J}_{ \omega}^{nl} + \mu_0
\epsilon_0 \epsilon_r (\partial \vec{E}_{SMPs}/\partial t)$, 
we obtain the wave equation as
\vspace{-0.25cm}
\begin{eqnarray}\nonumber
	\vec{\nabla}^2 \vec{E}_{SMPs} - \vec{\nabla} (\vec{\nabla} \cdot \vec{E}_{SMPs}) -
	\frac{\omega^2}{c^2} \left( \textrm{}
	\overline{\overline{\epsilon}} \vec{E}_{SMPs} \right) \\
	= -\mu_{0} i \omega \vec{J}_{\omega}^{nl} h \delta (x).
	\label{eq7}
	\vspace{-0.25cm}
\end{eqnarray}
Here, the components of effective permittivity  tensor $\overline{\overline{\epsilon}}$ for n-InSb at $\omega$ frequency given as: 
$\epsilon_{xx}=\epsilon_{zz}=\epsilon_{r}-(\epsilon_{r}\omega_{p}^{2} 
(\omega+i\nu))/(\omega {(\omega+i\nu)^{2}-\omega_{ce}^{2}})$, 
$\epsilon_{xz}=-\epsilon_{zx}=-i (\epsilon_{r}(\omega_{p}^{2} 
\omega_{ce})/(\omega( 
{(\omega+i\nu)^{2}-\omega_{ce}^{2}})))$, $\epsilon_{yy}=\epsilon_{r}-(\epsilon_{r
}\omega_{p}^{2})/(\omega {(\omega+i\nu)^{2}})$ and 
$\epsilon_{xy}=\epsilon_{yx}=\epsilon_{yz}=\epsilon_{zy}=0$. The delta function $\delta(x)$ models the $\vec{J}_{\omega}^{nl}$ confined to the rippled region.
Equation~\eqref{eq7} yields the expression
\vspace{-0.25cm}
\begin{multline}
	\frac{\partial^{2}E^z_{SMPs}}{\partial x^{2}} - \biggl[ k_z^2 {-
		\frac{\omega^2}{c^2} \biggl( \frac{\epsilon^2_{x z} + \epsilon^2_{x
				x}}{{\epsilon_{x x}}} \biggr)} \biggr] E^z_{SMPs} \\ = -
	\frac{c^2}{\epsilon_{x x} \omega^2} \biggl[ \biggl(
	{{\frac{\omega^2}{c^2} \epsilon_{x x}}} - k_z^2
	\biggr) J_{\omega, z}^{n l} \\ + \biggl(
	{{\frac{\omega^2}{c^2} \epsilon_{x z}}} + i k_z
	\partial_x \biggr)J_{\omega, x}^{n l} \biggr]  \mu_0 i \omega h\delta(x).
	\label{eq8}
	\vspace{-0.25cm}
\end{multline}
In the absence of the r.h.s term in Eq.~\eqref{eq8}, the solution reduces to the form
\vspace{-0.25cm}
\begin{equation}
	\vec{E}_{SMPs} = E_{SMPs} \vec{\psi} (x) e^{- i (\omega t - k_z z)}\label{eq9}
\end{equation}
here,
\vspace{-0.25cm}
\begin{equation}
	\vec{\psi}_(x) =
	\begin{cases}
		(\beta_{1 \omega}  \hat{x} +\hat{z}) e^{- \alpha_{1
				\omega} x},  \,\, \textrm {for air x} > 0 \\
		( \beta_{2 \omega}  \hat{x} + \hat{z}) e^{\alpha_{2
				\omega} x}, \,\,\,\,\textrm{for Graphene-n-InSb x} \leq 0 ;
	\end{cases}\nonumber
	\vspace{-0.15cm}
\end{equation}
\noindent
$\beta_{1 \omega}=-(i k_{z}/ \alpha_{1 \omega})$, $\alpha_{1 \omega}^2 = k_z^2 
-
(\omega^2/c^2)$, $\beta_{2 \omega} = \left( {\epsilon_{x z} \alpha_{2 \omega}+ \epsilon_{x x} i k_{z}}\right)/\left({ -\epsilon_{x x} \alpha_{2 \omega} + \epsilon_{x z} i k_{z} }\right) $ and $\alpha_{2 \omega}^2 = k_z^2 -
(\omega^2/c^2) (({\epsilon^2_{x z} + \epsilon^2_{x x}})/{\epsilon_{x x}})$.
%
The associated magnetic field of THz SMPs can be
written as
\begin{equation}
	{\vec{H}_{SMPs} = \hat{y} E_{SMPs}  e^{- i (\omega  t - k_z z)}
		\left\{\begin{array}{ll}
			i \frac{\omega \epsilon_0}{\alpha_{1\omega}} e^{- \alpha_{1\omega} x}, & \textrm{
				x} > 0\\
			\frac{i \epsilon_0 \omega \epsilon_{eff} \epsilon_{x x}}{-
\epsilon_{x x} {\alpha_{2\omega}}  + \epsilon_{x z} {i k_z} } e^{{\alpha }_{2\omega} x},&
		\textrm{ x} \leq 0\label{eq10}
		\end{array}\right.}
\end{equation}
The boundary condition $\vec{H}_{2, y} - \vec{H}_{1, y} =
\vec{J}_{\sigma_g}$ at $x = 0$, with $\vec{J}_{\sigma_g} = \sigma_g
\vec{E}_z$; yields the dispersion relation corresponding to THz SMPs as~\cite{liu2015directional}
\vspace{-0.25cm}
\begin{equation}
	\frac{\epsilon_{eff} \epsilon_{x x}}{\epsilon_{x x} {\alpha_{2\omega}}  -
		\epsilon_{x z} {i k_z} } + \frac{1}{\alpha_{1\omega}} = \frac{\sigma_g}{i \omega
		\epsilon_0}\label{eq11}
\end{equation}
where, $\epsilon_{eff} \!=\! (\epsilon_{xx}^{2}+\epsilon_{xz}^{2})/(\epsilon_{xx})$ and $\sigma_{g} \!=\! (i {\mathrm{e}}^2 \text{E}_\text{F}) / (\pi \hbar^2
	(\omega + i \nu))$.
%
For the phase matching, we require the condition 
\vspace{-0.25cm}
\begin{equation}
	q = k_{z} -k_{1z} + k_{2z}\label{eq12}
	\vspace{-0.5cm}
\end{equation}
\begin{figure}
  \centering
  \includegraphics[width=0.45\textwidth]{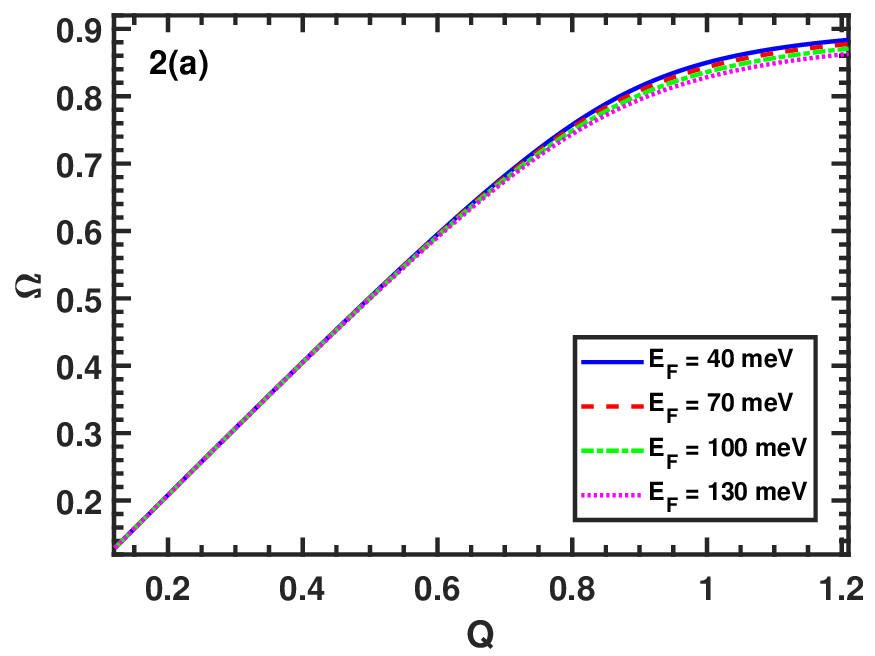}
  \includegraphics[width=0.45\textwidth]{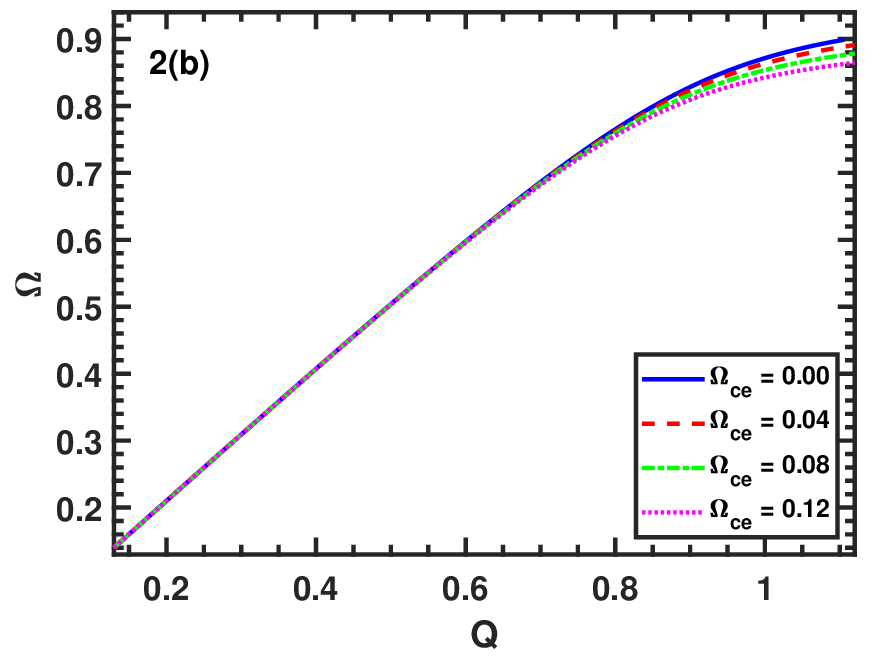}
  \includegraphics[width=0.45\textwidth]{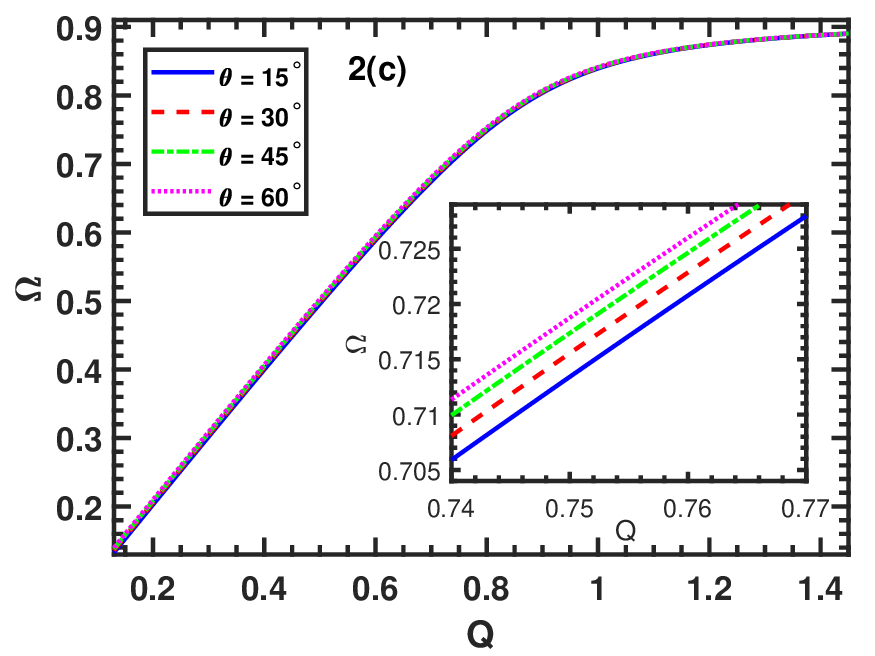}
\vspace{-0.25cm}
  \caption{Variation of $Q$ with $\Omega$: (a) for $\textrm{E}_\textrm{F} = 40  \textrm{ meV}, 70 \textrm{ meV},100 \textrm{ meV} $ and $130$ \textrm{ meV}, keeping $\Omega_{ce} = 0.12 $ and $\theta = 30^{\circ}$ constant, (b) for normalized magnetic field $\Omega_{ce} = 0, 0.04, 0.08$ and 0.12, keeping  $\theta = 30^{\circ}$ and $\textrm{E}_\textrm{F} = 110  \textrm{ meV}$ constant and (c) for incidence angle $\theta = 15^{\circ}$ to $60^{\circ}$, keeping $\textrm{E}_\textrm{F} = 110 \textrm{ meV}$ and $\Omega_{ce} = 0.12 $ constant.}
\label{Fig2}
\vspace{-0.6cm}
\end{figure}    

The wave number $q$ of the ripple supports the excitation of SMPs by providing the necessary momentum. Figure~\ref{Fig2}a shows the normalized ripple wave number $Q (= qc/\omega_{p})$
versus normalized THz frequency $\Omega$ ($=\omega/\omega_{p}$) for  $\text{E}_\text{F} = 20-130$ meV, keeping $\theta=30^{\circ}$ and $\Omega_{ce}$ ($=\omega_{ce}/\omega_{p}$)$ = 0.12$ constant.
In a similar way, Figure~\ref{Fig2}b depicts, $Q$ versus $\Omega$ for $\Omega_{ce} = 0.0 -0.12 $, keeping $\theta=30^{\circ}$ and $\text{E}_\text{F} = 110$ meV constant. Figure~\ref{Fig2}c presents $Q$ versus $\Omega$ for $\theta = 15^{\circ}$ to $60^{\circ}$, keeping $\Omega_{ce} = 0.12$ and $\text{E}_\text{F}$ = $110$ meV constant.
The parameters adopted here are identical to those used earlier for SMPs generation in laser-driven graphene–n-InSb systems~\cite{srivastav2025terahertz}. Since $q\propto k_z$, the dispersion results in 
Figs.\ref{Fig2}a, \ref{Fig2}b may appear to be similar to those
respective results of dispersion ($\Omega$ \textrm{vs} Q) reported by Srivastav 
and Kundu\cite{srivastav2025terahertz}. However, there are significant 
quantitative difference between results in Figs.\ref{Fig2}a, \ref{Fig2}b with
beating and respective results in Ref.\cite{srivastav2025terahertz} with a 
single laser, particularly at lower values of $Q < 0.6$. Thus present work 
envisages lower frequency $\Omega$ generation when compared to dispersion 
results in Ref.\cite{srivastav2025terahertz}.
A further characteristic of the dispersion emerges from the angular dependence depicted in the Figure~\ref{Fig2}c. For more clarity,
a comparison of the respective dispersion curves Figs.\ref{Fig2}a,
\ref{Fig2}b, \ref{Fig2}c along with those in Ref.\cite{srivastav2025terahertz} 
is shown in the Appendix. These differences in the dispersion characteristics 
are expected to yield
significant enhancement for the THz field amplitude as well.
Assuming that the THz SMPs retain the same mode structure, and only their amplitude varies along the 
$\hat{z}$ direction, we propose a solution of the following form
%
\begin{equation}
	\vec{E}_{SMPs} = E_{SMPs} (z) \vec{\psi} (x) e^{- i (\omega t - k_z z)}.
	\label{eq13}
\end{equation}
Solving Eq.\eqref{eq9} with Eq.\eqref{eq13} and transforming $k_z \rightarrow (k_z - i \partial / \partial z)$, we get
\vspace{-0.25cm}
\begin{multline}
	2 k_z \vec{\psi} (x) \frac{\partial E_{SMPs} (z)}{\partial z} e^{- i
		(\omega t - k_z z)}  = - \frac{c^2 \mu_0}{\omega \epsilon_{x x}} h \biggl[ \\
	{{\left( {\frac{\omega^2}{c^2}
				\epsilon_{x x}} - k_z^2 \right)}} J_{\omega, z}^{n l} \hat{z}  + \left(
	{{{\frac{\omega^2}{c^2} \epsilon_{x
					z}}}} + i k_z \alpha_{2 \omega} \right)  J_{\omega, x}^{n l} \hat{x}
	\biggr].\label{eq14}
\end{multline}
%
Multiplying \eqref{eq14} by $\vec{\psi}^{\ast}(x) \, dx$ and integrating $(-\infty, \infty)$, yields
%
\begin{multline}
	2 k_z \frac{\partial}{\partial z} (E_{SMPs}e^{- i (\omega t - k_z z)}) = -
	\frac{c^2 \mu_0}{\omega \epsilon_{x x}} h \biggl[ \\ \left(
	{{\frac{\omega^2}{c^2} \epsilon_{x x}}} - k_z^2
	\right) \frac{I_2}{I_1}  + \left(
	{{\frac{\omega^2}{c^2} \epsilon_{x z}}} + i k_z
	\alpha_{2 \omega} \right) \frac{I_3}{I_1} \biggr]\label{eq15}
\end{multline}
where, $ I_1 = \int^{\infty}_{- \infty} \vec{\psi} (x) \cdot
\vec{\psi}^{\ast} (x) d x $, $ I_2 = \int^{\infty}_{- \infty} \vec{\psi}^{\ast} (x) \cdot J_{\omega,
	z}^{n l} \hat{z} d x $ and $ I_3 = \int^{\infty}_{- \infty} \vec{\psi}^{\ast} (x) \cdot J_{\omega,
	x}^{n l} \hat{x} d x $.
By integrating Eq.\eqref{eq15} over the illumination length $d$, the expression for the field amplitude of THz SMPs is obtained as
\begin{multline}
	\biggl|\frac{E_{SMPs}}{E_1}\biggr| = \biggl|\frac{\omega^2_{p} E^{\ast}_2e d h }{4 \omega
		\epsilon_{x x} m_e^{\ast} k_{z}} \biggl( \frac{1 + \beta^2_{1 \omega}
	}{\alpha_{1 \omega}} + \frac{1 + \beta^2_{2 \omega} }{\alpha_{2
			\omega}} \biggr)^{-1}  \times \\ \biggl[ 2 \biggl(
	{\frac{\omega^2}{c^2} \epsilon_{x x}} - k_z^2
	\biggr) \tilde{J}_{\omega}^z  + \biggl(
	{\frac{\omega^2}{c^2} \epsilon_{x z}} + i k_z
	\alpha_{2 \omega} \biggr) \\ \biggl(\beta_{1 \omega} +
	\beta_{2 \omega}\biggr)\tilde{J}_{\omega}^x \biggr] \times T_{1,tr} T_{2,tr} \biggr| \label{eq16}
	\vspace{-0.25cm}
\end{multline}

Equation~\eqref{eq16} defines the normalized field amplitude (NFA) of THz SMPs, i.e., $|E_{SMPs}/E_{1}|$. It varies linearly with the transmission coefficients $T_{1,tr}$ and $T_{2,tr}$, as well as with the ripple height $h$ and the illumination length~$d$. The results are discussed in Sec.~\ref{sec4}.

\begin{figure}
	\centering
	\includegraphics[width=0.45\textwidth]{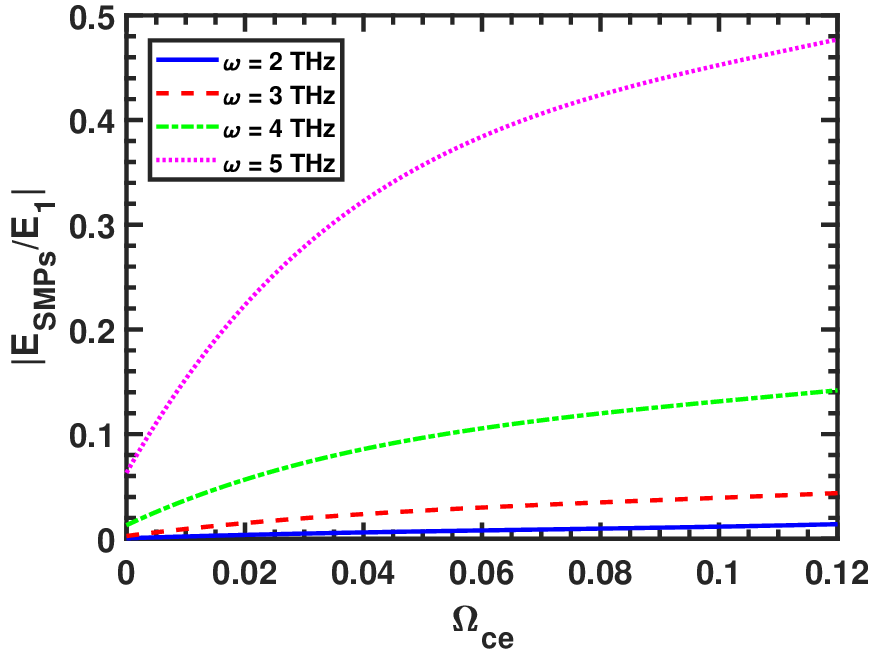} 
	\vspace{-0.25cm}		
	\caption{\label{Fig3} NFA of THz SMPs $|E_{SMPs}/E_{1}|$ versus $\Omega_{ce}$ for $\omega$ = 2 to 3 THz, with $\theta = 82.83^{\circ}$ and $\textrm{E}_\textrm{F}$= $120$ meV. Other parameters are  $\textrm{d} = 10 \mu m$, $\textrm{h} = 10 \mu m$, and  $\omega_{p} = 9.38$ THz.}
	\vspace{-0.25cm}			
\end{figure}
\vspace{-0.35cm}
\begin{figure}
	\centering
	\includegraphics[width=0.45\textwidth,height=0.35\textwidth]{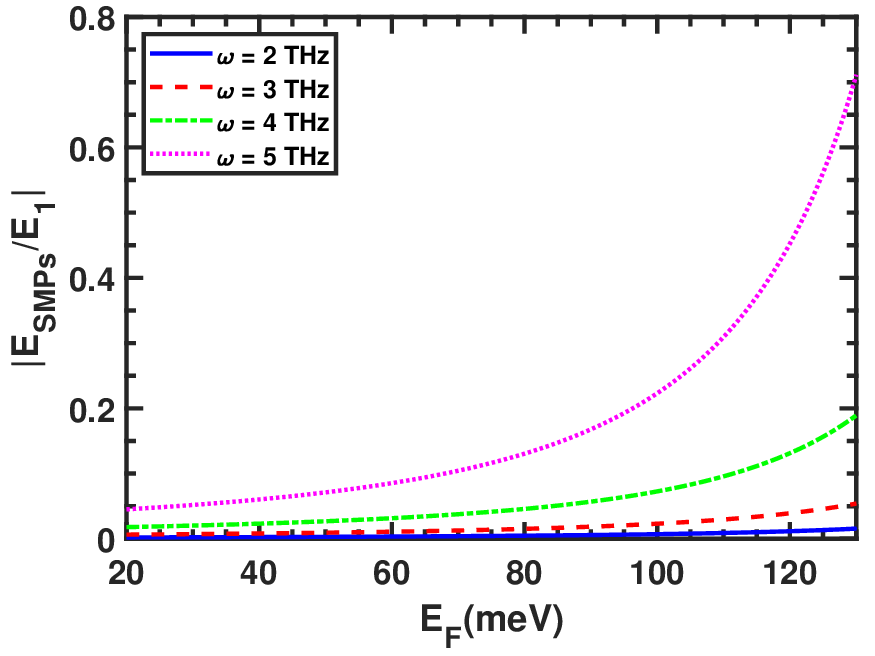} 
	\vspace{-0.25cm}		
	\caption{\label{Fig4} NFA of THz SMPs $|E_{SMPs}/E_{1}|$ versus $\text{E}_\text{F}$ for $\omega$ = 2 to 3 THz, with $\theta \!=\! 82.83^{\circ}$ and  $\Omega_{ce}\!=\! 0.04$ constant. All other parameters are consistent with those in Fig.~\ref{Fig3}.}
	\vspace{-0.25cm}
\end{figure}
\vspace{-0.35cm}
\begin{figure}
	\centering
	\includegraphics[width=0.45\textwidth,height=0.35\textwidth]{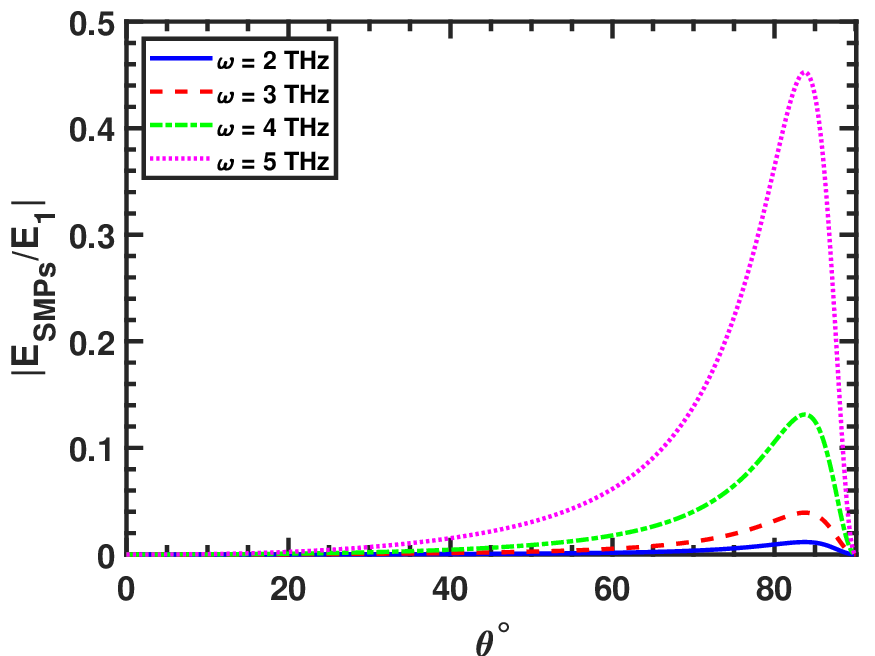} 
	\vspace{-0.25cm}		
	\caption{\label{Fig5} NFA of THz SMPs $|E_{SMPs}/E_{1}|$ versus $\theta^{\circ}$  for THz frequency $\omega$ = 2 to 3 THz, keeping $\Omega_{ce}= 0.04$ and $\textrm{E}_\textrm{F}$ = $120$ meV constant. 
	All remaining parameters are consistent with those in Fig.\ref{Fig3}.}
	\vspace{-0.25cm}
\end{figure}

\begin{figure}
	\centering
	\includegraphics[width=0.45\textwidth,height=0.325\textwidth]{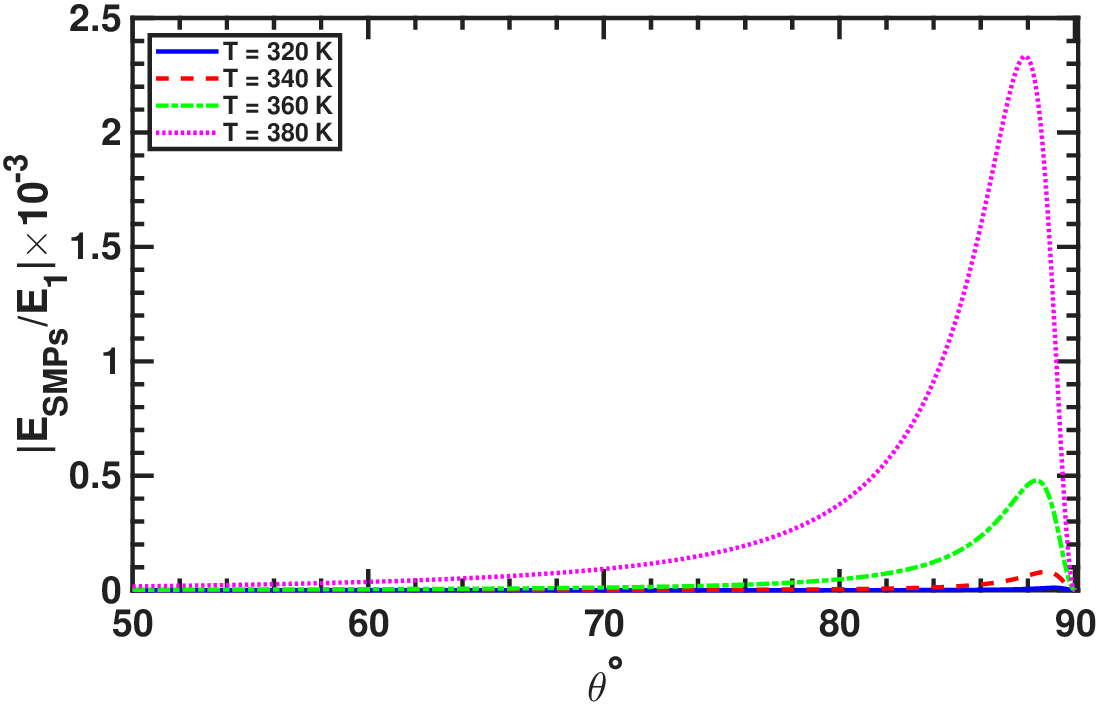} 
	\vspace{-0.25cm}		
	\caption{\label{Fig6} NFA of THz SMPs $|E_{SMPs}/E_{1}|$ versus  $\theta^{\circ}$ for n-type semiconductor temperature T $=300 \textrm { to } 360$ K, keeping $\textrm{E}_\textrm{F}$ = $120$ meV, $\omega = 5 $THz and $\theta = 82.83^{\circ}$ constant. All remaining parameters are consistent with those in Fig.~\ref{Fig5}.}
	\vspace{-0.25cm}
\end{figure}

	\section{Results and Discussion}\label{sec4} 
\vspace{-0.35cm}
To investigate the behavior of the NFA of THz SMPs, represented as $\vert E_{\text{SMPs}}/E_1 \vert$, we consider the interaction of two $\mathrm{CO}_2$ laser beams with wavelengths $\lambda_1 = 10.64\ \mu\text{m}$ and $\lambda_2 = 10.81\ \mu\text{m}$, and same intensity of $I = 2 \times 10^{15}\ \text{W/cm}^2$. The analysis is conducted under collisionless conditions (i.e., $\nu = 0$). The illumination length and surface ripple height are both fixed at $d = h = 10\ \mu\text{m}$. The underlying substrate is n-type indium antimonide (n-InSb), characterized by a relative permittivity of $\epsilon_r = 15.68$, an electron density of $n_0 = 2.4 \times 10^{23}\ \text{m}^{-3}$, and an electron plasma frequency of $\omega_p = 9.38$ THz. In this study, the normalized THz frequency $\Omega$ is varied in the range 0.2 to 0.5, while the $\Omega_{ce}$ is varied from 0 to 0.12. Additionally, the graphene sheet's  Fermi energy is varied between 20 to 130 meV to explore its tunability on THz SMPs excitation. These parameter values are consistent with those reported in earlier investigations on laser-induced excitation of SPs and SMPs in graphene and semiconductor systems\cite{Srivastav2022,10.1117/1.JNP.11.036015}.

\vspace{-0.5cm}
\subsection{\textbf{Competing Effects of Magnetic Field and Fermi Energy on THz SMP Amplitude}}\label{sec4.3}
\vspace{-0.35cm}
Figure~\ref{Fig3} represents NFA of THz SMPs $\vert E_{SMPs}/E_{1}\vert$ versus $\Omega_{ce}$ for $\omega$ = 2 to 5 THz, keeping $\textrm{E}_\textrm{F} = 120$ meV and $\theta^{\circ} = 82.83^{\circ}$ constant. The NFA of THz SMPs $\vert E_{\text{SMPs}}/E_1 \vert$ is found to grow as $\Omega_{ce}$ increases. This behavior arises from the increasingly constrained motion of electrons as the externally applied static magnetic field becomes stronger~\cite{srivastav2025terahertz}. Figure.~\ref{Fig4} shows
$\vert E_{SMPs}/E_{1}\vert$ versus 
$\textrm{E}_\textrm{F}$ for $\omega$ = 2 to 5 THz, keeping $\Omega_{ce}=0.04$ and
$\theta = 82.83^{\circ}$ constant. For each $\omega$, $\vert E_{SMPs}/E_{1}\vert$ increases as $\textrm{E}_\textrm{F}$ increases. The NFA of THz SMPs wave is found to increase with both the $\text{E}_\text{F}$ and  $B_{0}$, as illustrated in Figs.~\ref{Fig3} and \ref{Fig4}. This trend is driven by the rise in the $k_z$, propagation constant, observed in the dispersion characteristics of THz SMPs figures.\cite{srivastav2025terahertz} with a rise in magnetic field strength and Fermi energy. that is $\vert E_{SMPs}/E_{1}\vert\propto k_{z}^{2}$
indicating that the NFA of THz SMPs grows with the normalized propagation constant $K_{z}$.

%
%


\vspace{-0.5cm}
\subsection{\textbf{Angular Dependence and Resonant Excitation}}\label{sec4.5}
\vspace{-0.25cm}
Figure~\ref{Fig5} shows $\vert E_{SMPs}/E_{1}\vert$ versus incident angle $\theta$ for $\omega$ = 2 to 5 THz, keeping $\Omega_{ce}=0.04$ and $\textrm{E}_\textrm{F} = 120$ meV constant. The NFA of THz SMPs field, $\vert E_{SMPs}/E_{1}\vert$, initially rises with increasing $\theta^{\circ}$, reaches a peak near grazing incidence ($\theta^{\circ}\rightarrow
90^{\circ}$) depending on $\omega$, and subsequently decreases for larger $\theta^{\circ}$. This shows that the effects are most prominent around $\theta \approx 83^{\circ}$ due to resonance. At normal incidence no THz is induced. Accordingly, an angle $\theta = 82.83^{\circ}$ is used in Figs.~\ref{Fig3} and \ref{Fig4} to maximize result. At the normalized THz frequency corresponding to SMPs resonance, the NFA of THz SMPs wave attains its maximum amplitude (Figure~\ref{Fig5})~\cite{srivastav2025terahertz}.

\vspace{-0.5cm}
\subsection{\textbf{Frequency and Thermal Tunability}}\label{sec4.6}  
\vspace{-0.25cm}
   Figures~\ref{Fig3} to \ref{Fig5} demonstrate that the NFA of THz SMPs wave grows with the THz frequency $\omega$. This trend can be attributed to the rise in the required normalized $Q$, which itself grows with the THz frequency $\Omega$, as depicted in Figure~\ref{Fig2}.

    Figure~\ref{Fig6} shows $\vert E_{SMPs}/E_{1}\vert$ versus $\theta^{\circ}$  for n-InSb semiconductor temperature T=320 to 380 K, keeping $\textrm{E}_\textrm{F} = 120$ meV, $\omega=5$ THz and $\Omega_{ce} = 0.12$ constant.
   The NFA $\vert E_{SMPs}/E_{1}\vert$ increases with the n-InSb's temperature. This behavior arises from the temperature-dependent increase in electron density $n_0$, which enhances the plasma frequency according to $\omega_p \propto \sqrt{n_0} \propto \sqrt{T^{3/2} \exp(-{E_{g}}/2k_B T)}$\cite{jing2022thermally, gao2023multifunctional}. Consequently, $\vert E_{\text{SMPs}}/E_1 \vert$ also increases, which is consistent with Eq.\eqref{eq16}; as $\vert E_{\text{SMPs}}/E_1 \vert \propto \omega_p^2 \propto T^{3/2} \exp(-{E_{g}}/2k_B T)$.

   \vspace{-0.35cm}
	\section{Conclusion}\label{sec5}
    \vspace{-0.35cm}
	In the present work, we have analytically investigated the influence of key physical parameters: including the semiconductor temperature $\text{T}$, externally applied static magnetic field strength $\text{B}_{0}$, Fermi energy of graphene $\text{E}_\text{F}$, and the incidence angle $\theta$ of p-polarized laser beams on the amplitude of THz wave generated via the beating of two p-polarized laser frequencies. The analysis focuses on the feasibility of THz SMPs excitation within the frequency range of approximately 2 THz to 5 THz, while systematically varying the $\textrm{E}_\textrm{F} = 20-130$~meV and strength of $B_{0}\approx 0 - 0.09$~Tesla.

 In the earlier work we investigated THz wave generation by {\emph{single}} $\mathrm{CO}_2$
laser\cite{srivastav2025terahertz} and reported NFA of THz SMPs of the 
order of $10^{-2}$. However, here we report (as in Figs.~\ref{Fig3}-\ref{Fig5}) NFA of THz SMPs of the order of $10^{-1}$ using {\emph{beating of two}} $\mathrm{CO}_2$ lasers. The field amplitude is now one-order
higher, and increases gradually in contrast. Notably, the maximum
normalized THz SMPs field amplitude $\vert E_{\text{SMPs}}/E_1 \vert\sim 10^{-1}$ (as illustrated in
Figs.~\ref{Fig3}-\ref{Fig5}) represents \emph{nearly a two
order} of magnitude enhancement over the previously reported values for the similar
excitation mechanisms\cite{Srivastav2022,verma2021terahertz}. This significant 
increase in the field THz field amplitude highlights the efficiency and potential of the proposed scheme for
high-performance THz wave generation. The results also indicate that a higher
magnetic field, combined with an increase in temperature, enhances the THz field 
intensity.

The present findings open new possibilities for the development of advanced THz
technologies, with potential applications in active plasmonic devices, 
high-sensitivity sensors, ultrafast detection, and re-configurable 
optoelectronic 
systems\cite{son2019potential,Angelina2021,salameh20225g,kaur2024highly}.

\vspace{-0.35cm}
\section{Appendix: Comparison of dispersion characteristics between NLM and LMC}
\label{Appendix}
\begin{figure}[h]
\vspace{-0.25cm}
	\centering
	
	\includegraphics[width=0.425\textwidth]{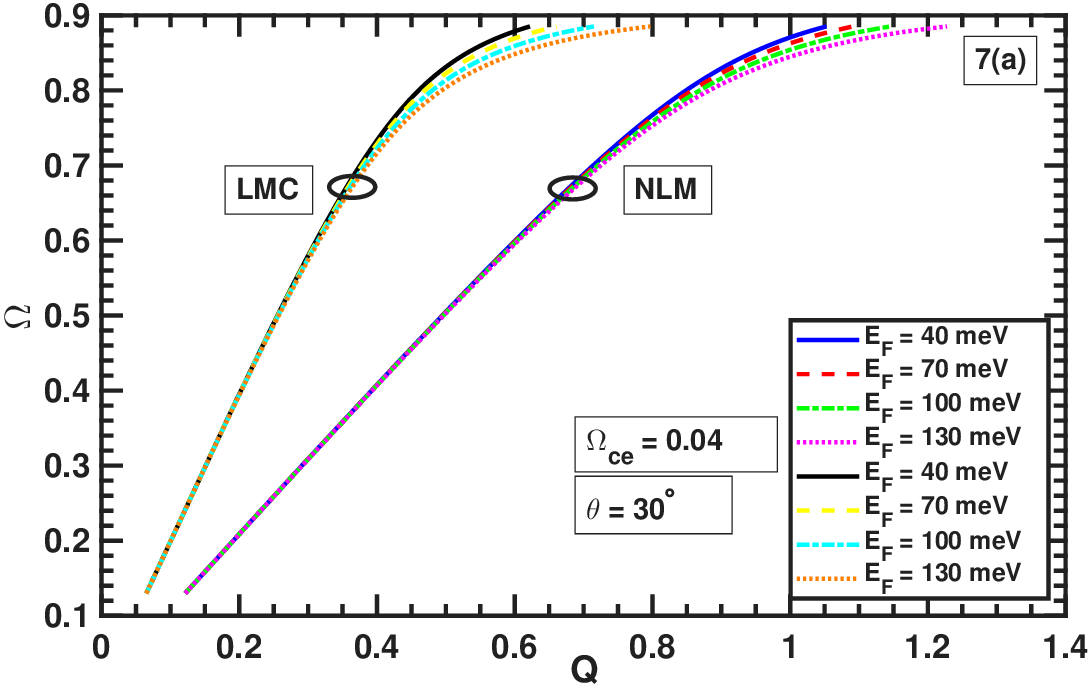}
	\includegraphics[width=0.425\textwidth]{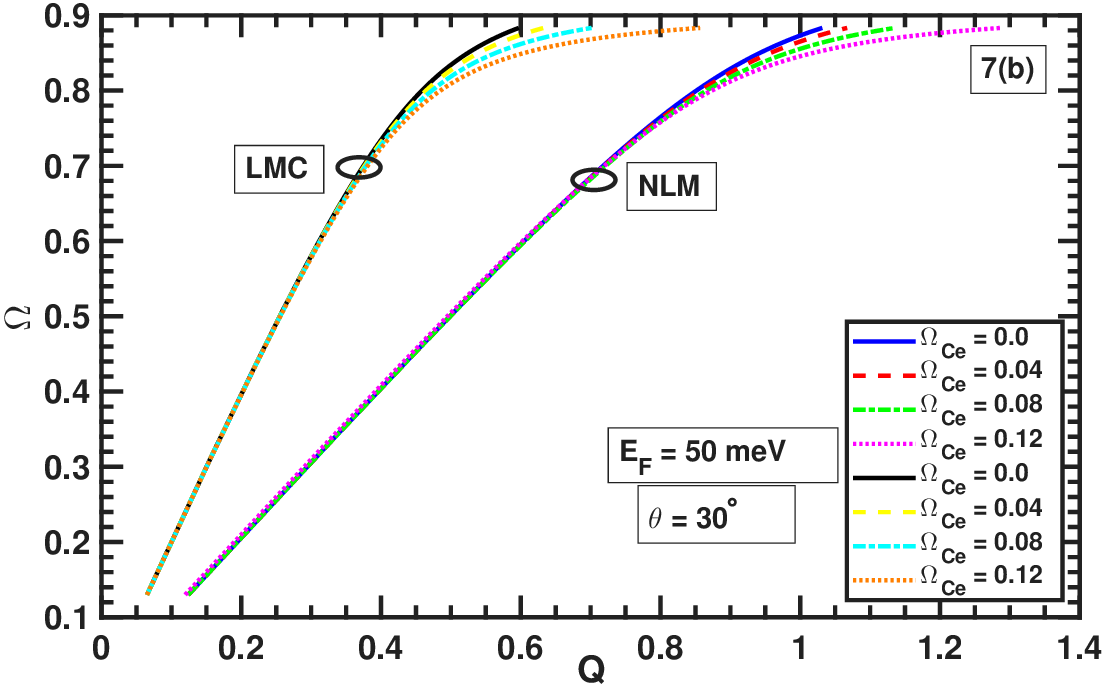}
	\includegraphics[width=0.425\textwidth]{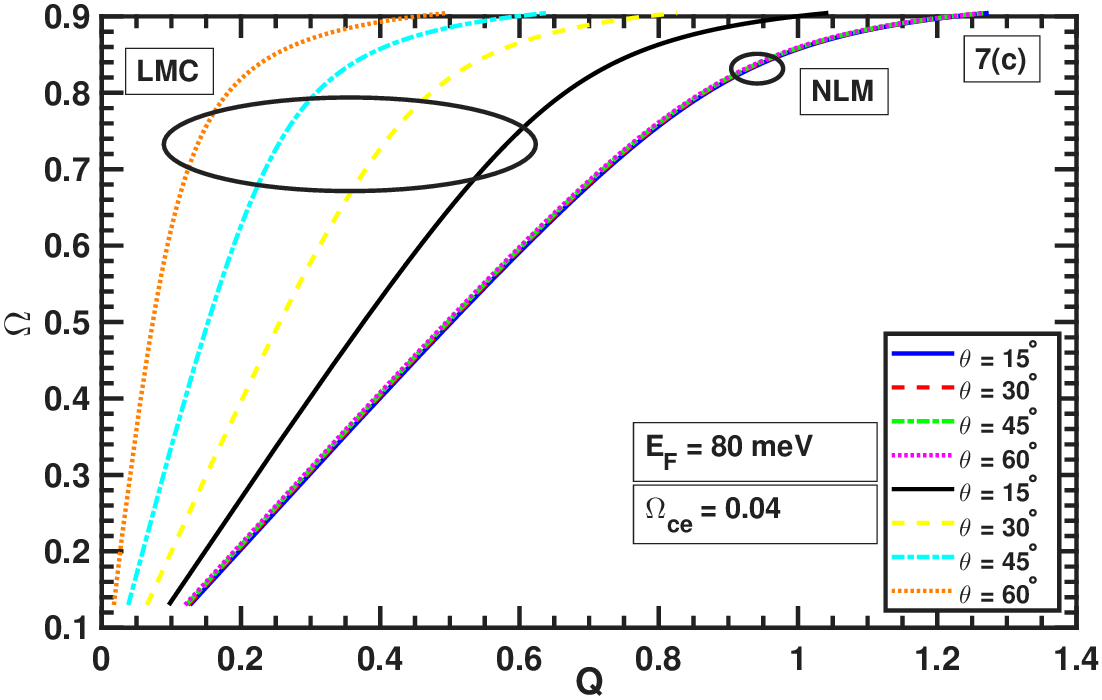}
\vspace{-0.25cm}
	\caption{Variation of normalized ripple wave number $Q$ with normalized THz 
frequency $\Omega$: (a) for  
$\textrm{E}_\textrm{F} = 40  \textrm{ meV}, 70 \textrm{ meV},100 \textrm{ meV} 
$ 
and $130$ \textrm{ meV}, keeping $\Omega_{ce} = 0.04 $ and $\theta = 
30^{\circ}$ constant, (b) for $\Omega_{ce} = 0, 
0.04, 0.08$ and 0.12, keeping $\textrm{E}_\textrm{F} = 50  \textrm{ 
meV}$ 
and $\theta = 30^{\circ}$ constant and (c) for incidence angle $\theta = 
15^{\circ}$ to $60^{\circ}$, keeping 
$\textrm{E}_\textrm{F} = 80 \textrm{ meV}$ and $\Omega_{ce} = 0.04 $ constant. NLM 
curves are the dispersion curves in the present 
work (Figs.~\ref{Fig2}a,\ref{Fig2}b,\ref{Fig2}c), while LMC curves correspond
to the dispersion curves from Ref.\cite{srivastav2025terahertz}.}
	\label{Fig7}
\end{figure}

\vspace{-0.25cm}
Equation \eqref{eq12} represents the phase-matching condition for the resonant 
excitation of THz SMPs wave via NLM of two lasers, and 
corresponding dispersion curves are shown in 
Figs.~\ref{Fig2}a,\ref{Fig2}b,\ref{Fig2}c. In contrast, $q = k_{z}-k_{0z}$
corresponds to the phase-matching condition for the resonant laser-excitation 
of 
THz SMPs through LMC, as reported by Srivastav and 
Kundu\cite{srivastav2025terahertz} and the corresponding dispersion curves are 
reproduced in Figs.~\ref{Fig7}a,\ref{Fig7}b,\ref{Fig7}c for comparison with
same parameters. There are significant 
quantitative difference between NLM results 
with 
beating and respective LMC results with a 
single laser, particularly at lower values of $Q < 0.6$. The dispersion results 
in the present work 
envisages lower frequency $\Omega$ generation compared to dispersion 
results in Ref.\cite{srivastav2025terahertz} for same parameters.
Additionally, the dispersion characteristics may yield
significant enhancement for the THz field amplitude as well.

\vspace{-0.35cm}
\section{Author Declarations}
\vspace{-0.35cm}
\subsection{Conflict of interest} 
\vspace{-0.5cm}
The authors have no conflicts to disclose.

\vspace{-0.5cm}
\subsection{Author Contributions} 
\vspace{-0.35cm}
Rohit Kumar Srivastav (RKS) and Mrityunjay Kundu (MK) carried out the research. RKS prepared the figures and drafted the original manuscript. MK reviewed and edited the final version.

\vspace{-0.35cm}
\section{Data Availability}
\vspace{-0.35cm}
The data that supports the findings of this study are available from the corresponding author upon reasonable request.

\section{ORCID IDs}
\vspace{-0.35cm}
\noindent
Rohit Kumar Srivartav: {https://orcid.org/0000-0002-3494-1218}\\
Mrityunjay Kundu: {https://orcid.org/0009-0002-5221-5639}

\section*{References}
\bibliographystyle{aipnum4-2}
\bibliography{Reference}

\providecommand{\noopsort}[1]{}\providecommand{\singleletter}[1]{#1}%
\begin{thebibliography}{50}%
\makeatletter
\providecommand \@ifxundefined [1]{%
 \@ifx{#1\undefined}
}%
\providecommand \@ifnum [1]{%
 \ifnum #1\expandafter \@firstoftwo
 \else \expandafter \@secondoftwo
 \fi
}%
\providecommand \@ifx [1]{%
 \ifx #1\expandafter \@firstoftwo
 \else \expandafter \@secondoftwo
 \fi
}%
\providecommand \natexlab [1]{#1}%
\providecommand \enquote  [1]{``#1''}%
\providecommand \bibnamefont  [1]{#1}%
\providecommand \bibfnamefont [1]{#1}%
\providecommand \citenamefont [1]{#1}%
\providecommand \href@noop [0]{\@secondoftwo}%
\providecommand \href [0]{\begingroup \@sanitize@url \@href}%
\providecommand \@href[1]{\@@startlink{#1}\@@href}%
\providecommand \@@href[1]{\endgroup#1\@@endlink}%
\providecommand \@sanitize@url [0]{\catcode `\\12\catcode `\$12\catcode
  `\&12\catcode `\#12\catcode `\^12\catcode `\_12\catcode `\%12\relax}%
\providecommand \@@startlink[1]{}%
\providecommand \@@endlink[0]{}%
\providecommand \url  [0]{\begingroup\@sanitize@url \@url }%
\providecommand \@url [1]{\endgroup\@href {#1}{\urlprefix }}%
\providecommand \urlprefix  [0]{URL }%
\providecommand \Eprint [0]{\href }%
\providecommand \doibase [0]{https://doi.org/}%
\providecommand \selectlanguage [0]{\@gobble}%
\providecommand \bibinfo  [0]{\@secondoftwo}%
\providecommand \bibfield  [0]{\@secondoftwo}%
\providecommand \translation [1]{[#1]}%
\providecommand \BibitemOpen [0]{}%
\providecommand \bibitemStop [0]{}%
\providecommand \bibitemNoStop [0]{.\EOS\space}%
\providecommand \EOS [0]{\spacefactor3000\relax}%
\providecommand \BibitemShut  [1]{\csname bibitem#1\endcsname}%
\let\auto@bib@innerbib\@empty
\bibitem [{\citenamefont {Liu}\ \emph {et~al.}(2024{\natexlab{a}})\citenamefont
  {Liu}, \citenamefont {Wang}, \citenamefont {Sun}, \citenamefont {Zhu},\ and\
  \citenamefont {Yin}}]{liu2024electronic}%
  \BibitemOpen
  \bibfield  {author} {\bibinfo {author} {\bibfnamefont {Y.-Q.}\ \bibnamefont
  {Liu}}, \bibinfo {author} {\bibfnamefont {Y.}~\bibnamefont {Wang}}, \bibinfo
  {author} {\bibfnamefont {J.}~\bibnamefont {Sun}}, \bibinfo {author}
  {\bibfnamefont {Y.}~\bibnamefont {Zhu}},\ and\ \bibinfo {author}
  {\bibfnamefont {H.}~\bibnamefont {Yin}},\ }\href@noop {} {\bibfield
  {journal} {\bibinfo  {journal} {Physics of Plasmas}\ }\textbf {\bibinfo
  {volume} {31}} (\bibinfo {year} {2024}{\natexlab{a}})}\BibitemShut {NoStop}%
\bibitem [{\citenamefont {Nenninger}, \citenamefont {Piliarik},\ and\
  \citenamefont {Homola}(2002)}]{nenninger2002data}%
  \BibitemOpen
  \bibfield  {author} {\bibinfo {author} {\bibfnamefont {G.~G.}\ \bibnamefont
  {Nenninger}}, \bibinfo {author} {\bibfnamefont {M.}~\bibnamefont
  {Piliarik}},\ and\ \bibinfo {author} {\bibfnamefont {J.}~\bibnamefont
  {Homola}},\ }\href@noop {} {\bibfield  {journal} {\bibinfo  {journal}
  {Measurement Science and Technology}\ }\textbf {\bibinfo {volume} {13}},\
  \bibinfo {pages} {2038} (\bibinfo {year} {2002})}\BibitemShut {NoStop}%
\bibitem [{\citenamefont {Masson}(2017)}]{masson2017surface}%
  \BibitemOpen
  \bibfield  {author} {\bibinfo {author} {\bibfnamefont {J.-F.}\ \bibnamefont
  {Masson}},\ }\href@noop {} {\bibfield  {journal} {\bibinfo  {journal} {ACS
  sensors}\ }\textbf {\bibinfo {volume} {2}},\ \bibinfo {pages} {16} (\bibinfo
  {year} {2017})}\BibitemShut {NoStop}%
\bibitem [{\citenamefont {Zhang}\ \emph {et~al.}(2023)\citenamefont {Zhang},
  \citenamefont {Xiao}, \citenamefont {Zhao}, \citenamefont {Liu},
  \citenamefont {Yu}, \citenamefont {Hu}, \citenamefont {Chu}, \citenamefont
  {Xu}, \citenamefont {Liu}, \citenamefont {Zou}, \citenamefont {Zhang},
  \citenamefont {Liu},\ and\ \citenamefont {Li}}]{10106756}%
  \BibitemOpen
  \bibfield  {author} {\bibinfo {author} {\bibfnamefont {C.}~\bibnamefont
  {Zhang}}, \bibinfo {author} {\bibfnamefont {P.}~\bibnamefont {Xiao}},
  \bibinfo {author} {\bibfnamefont {Z.-T.}\ \bibnamefont {Zhao}}, \bibinfo
  {author} {\bibfnamefont {Z.}~\bibnamefont {Liu}}, \bibinfo {author}
  {\bibfnamefont {J.}~\bibnamefont {Yu}}, \bibinfo {author} {\bibfnamefont
  {X.-Y.}\ \bibnamefont {Hu}}, \bibinfo {author} {\bibfnamefont {H.-B.}\
  \bibnamefont {Chu}}, \bibinfo {author} {\bibfnamefont {J.-J.}\ \bibnamefont
  {Xu}}, \bibinfo {author} {\bibfnamefont {M.-Y.}\ \bibnamefont {Liu}},
  \bibinfo {author} {\bibfnamefont {Q.}~\bibnamefont {Zou}}, \bibinfo {author}
  {\bibfnamefont {L.}~\bibnamefont {Zhang}}, \bibinfo {author} {\bibfnamefont
  {Q.}~\bibnamefont {Liu}},\ and\ \bibinfo {author} {\bibfnamefont {G.-S.}\
  \bibnamefont {Li}},\ }\href {https://doi.org/10.1109/JSEN.2023.3266262}
  {\bibfield  {journal} {\bibinfo  {journal} {IEEE Sensors Journal}\ }\textbf
  {\bibinfo {volume} {23}},\ \bibinfo {pages} {11591} (\bibinfo {year}
  {2023})}\BibitemShut {NoStop}%
\bibitem [{\citenamefont {Dong}, \citenamefont {Xu},\ and\ \citenamefont
  {Cheng}(2024)}]{dong2024enhanced}%
  \BibitemOpen
  \bibfield  {author} {\bibinfo {author} {\bibfnamefont {P.}~\bibnamefont
  {Dong}}, \bibinfo {author} {\bibfnamefont {J.}~\bibnamefont {Xu}},\ and\
  \bibinfo {author} {\bibfnamefont {J.}~\bibnamefont {Cheng}},\ }\href@noop {}
  {\bibfield  {journal} {\bibinfo  {journal} {Physics of Plasmas}\ }\textbf
  {\bibinfo {volume} {31}} (\bibinfo {year} {2024})}\BibitemShut {NoStop}%
\bibitem [{\citenamefont {Pitarke}\ \emph {et~al.}(2006)\citenamefont
  {Pitarke}, \citenamefont {Silkin}, \citenamefont {Chulkov},\ and\
  \citenamefont {Echenique}}]{pitarke2006theory}%
  \BibitemOpen
  \bibfield  {author} {\bibinfo {author} {\bibfnamefont {J.}~\bibnamefont
  {Pitarke}}, \bibinfo {author} {\bibfnamefont {V.}~\bibnamefont {Silkin}},
  \bibinfo {author} {\bibfnamefont {E.}~\bibnamefont {Chulkov}},\ and\ \bibinfo
  {author} {\bibfnamefont {P.}~\bibnamefont {Echenique}},\ }\href@noop {}
  {\bibfield  {journal} {\bibinfo  {journal} {Reports on progress in physics}\
  }\textbf {\bibinfo {volume} {70}},\ \bibinfo {pages} {1} (\bibinfo {year}
  {2006})}\BibitemShut {NoStop}%
\bibitem [{\citenamefont {Zhang}, \citenamefont {Zhang},\ and\ \citenamefont
  {Xu}(2012)}]{zhang2012surface}%
  \BibitemOpen
  \bibfield  {author} {\bibinfo {author} {\bibfnamefont {J.}~\bibnamefont
  {Zhang}}, \bibinfo {author} {\bibfnamefont {L.}~\bibnamefont {Zhang}},\ and\
  \bibinfo {author} {\bibfnamefont {W.}~\bibnamefont {Xu}},\ }\href@noop {}
  {\bibfield  {journal} {\bibinfo  {journal} {Journal of Physics D: Applied
  Physics}\ }\textbf {\bibinfo {volume} {45}},\ \bibinfo {pages} {113001}
  (\bibinfo {year} {2012})}\BibitemShut {NoStop}%
\bibitem [{\citenamefont {Han}\ and\ \citenamefont
  {Bozhevolnyi}(2012)}]{han2012radiation}%
  \BibitemOpen
  \bibfield  {author} {\bibinfo {author} {\bibfnamefont {Z.}~\bibnamefont
  {Han}}\ and\ \bibinfo {author} {\bibfnamefont {S.~I.}\ \bibnamefont
  {Bozhevolnyi}},\ }\href@noop {} {\bibfield  {journal} {\bibinfo  {journal}
  {Reports on Progress in Physics}\ }\textbf {\bibinfo {volume} {76}},\
  \bibinfo {pages} {016402} (\bibinfo {year} {2012})}\BibitemShut {NoStop}%
\bibitem [{\citenamefont {de~Hoogh}\ \emph {et~al.}(2016)\citenamefont
  {de~Hoogh}, \citenamefont {Opheij}, \citenamefont {Wulf}, \citenamefont
  {Rotenberg},\ and\ \citenamefont {Kuipers}}]{de2016harmonics}%
  \BibitemOpen
  \bibfield  {author} {\bibinfo {author} {\bibfnamefont {A.}~\bibnamefont
  {de~Hoogh}}, \bibinfo {author} {\bibfnamefont {A.}~\bibnamefont {Opheij}},
  \bibinfo {author} {\bibfnamefont {M.}~\bibnamefont {Wulf}}, \bibinfo {author}
  {\bibfnamefont {N.}~\bibnamefont {Rotenberg}},\ and\ \bibinfo {author}
  {\bibfnamefont {L.}~\bibnamefont {Kuipers}},\ }\href@noop {} {\bibfield
  {journal} {\bibinfo  {journal} {ACS photonics}\ }\textbf {\bibinfo {volume}
  {3}},\ \bibinfo {pages} {1446} (\bibinfo {year} {2016})}\BibitemShut
  {NoStop}%
\bibitem [{\citenamefont {Sedaghat~Nejad}\ and\ \citenamefont
  {Ghasempour~Ardakani}(2024)}]{sedaghat2024giant}%
  \BibitemOpen
  \bibfield  {author} {\bibinfo {author} {\bibfnamefont {M.}~\bibnamefont
  {Sedaghat~Nejad}}\ and\ \bibinfo {author} {\bibfnamefont {A.}~\bibnamefont
  {Ghasempour~Ardakani}},\ }\href@noop {} {\bibfield  {journal} {\bibinfo
  {journal} {Scientific Reports}\ }\textbf {\bibinfo {volume} {14}},\ \bibinfo
  {pages} {2853} (\bibinfo {year} {2024})}\BibitemShut {NoStop}%
\bibitem [{\citenamefont {Kumar}\ and\ \citenamefont
  {Tripathi}(2007)}]{kumar2007filamentation}%
  \BibitemOpen
  \bibfield  {author} {\bibinfo {author} {\bibfnamefont {G.}~\bibnamefont
  {Kumar}}\ and\ \bibinfo {author} {\bibfnamefont {V.}~\bibnamefont
  {Tripathi}},\ }\href@noop {} {\bibfield  {journal} {\bibinfo  {journal}
  {Journal of Applied Physics}\ }\textbf {\bibinfo {volume} {102}} (\bibinfo
  {year} {2007})}\BibitemShut {NoStop}%
\bibitem [{\citenamefont {Srivastav}\ and\ \citenamefont
  {Panwar}(2024)}]{srivastav2024resonant}%
  \BibitemOpen
  \bibfield  {author} {\bibinfo {author} {\bibfnamefont {R.~K.}\ \bibnamefont
  {Srivastav}}\ and\ \bibinfo {author} {\bibfnamefont {A.}~\bibnamefont
  {Panwar}},\ }\href@noop {} {\bibfield  {journal} {\bibinfo  {journal}
  {Journal of Plasma Physics}\ }\textbf {\bibinfo {volume} {90}},\ \bibinfo
  {pages} {905900106} (\bibinfo {year} {2024})}\BibitemShut {NoStop}%
\bibitem [{\citenamefont {Liu}\ \emph {et~al.}(2024{\natexlab{b}})\citenamefont
  {Liu}, \citenamefont {Zhang}, \citenamefont {Wang}, \citenamefont {Cai},
  \citenamefont {Sun}, \citenamefont {Zhu},\ and\ \citenamefont
  {Li}}]{liu2024excitation}%
  \BibitemOpen
  \bibfield  {author} {\bibinfo {author} {\bibfnamefont {Y.}~\bibnamefont
  {Liu}}, \bibinfo {author} {\bibfnamefont {X.}~\bibnamefont {Zhang}}, \bibinfo
  {author} {\bibfnamefont {Y.}~\bibnamefont {Wang}}, \bibinfo {author}
  {\bibfnamefont {H.}~\bibnamefont {Cai}}, \bibinfo {author} {\bibfnamefont
  {J.}~\bibnamefont {Sun}}, \bibinfo {author} {\bibfnamefont {Y.}~\bibnamefont
  {Zhu}},\ and\ \bibinfo {author} {\bibfnamefont {L.}~\bibnamefont {Li}},\
  }\href@noop {} {\bibfield  {journal} {\bibinfo  {journal} {Micromachines}\
  }\textbf {\bibinfo {volume} {15}},\ \bibinfo {pages} {293} (\bibinfo {year}
  {2024}{\natexlab{b}})}\BibitemShut {NoStop}%
\bibitem [{\citenamefont {Chamoli}, \citenamefont {Gupta},\ and\ \citenamefont
  {Kumar}(2024)}]{chamoli2024surface}%
  \BibitemOpen
  \bibfield  {author} {\bibinfo {author} {\bibfnamefont {A.}~\bibnamefont
  {Chamoli}}, \bibinfo {author} {\bibfnamefont {D.~N.}\ \bibnamefont {Gupta}},\
  and\ \bibinfo {author} {\bibfnamefont {V.}~\bibnamefont {Kumar}},\
  }\href@noop {} {\bibfield  {journal} {\bibinfo  {journal} {Radiation Effects
  and Defects in Solids}\ }\textbf {\bibinfo {volume} {179}},\ \bibinfo {pages}
  {962} (\bibinfo {year} {2024})}\BibitemShut {NoStop}%
\bibitem [{\citenamefont {Molavi~Choobini}\ and\ \citenamefont
  {Aghamir}(2025)}]{molavi2025role}%
  \BibitemOpen
  \bibfield  {author} {\bibinfo {author} {\bibfnamefont {A.~A.}\ \bibnamefont
  {Molavi~Choobini}}\ and\ \bibinfo {author} {\bibfnamefont {F.~M.}\
  \bibnamefont {Aghamir}},\ }\href@noop {} {\bibfield  {journal} {\bibinfo
  {journal} {Waves in Random and Complex Media}\ }\textbf {\bibinfo {volume}
  {35}},\ \bibinfo {pages} {2250} (\bibinfo {year} {2025})}\BibitemShut
  {NoStop}%
\bibitem [{\citenamefont {Mishra}\ \emph {et~al.}(2025)\citenamefont {Mishra},
  \citenamefont {Singh}, \citenamefont {Kumar},\ and\ \citenamefont
  {Jha}}]{mishra2025twisted}%
  \BibitemOpen
  \bibfield  {author} {\bibinfo {author} {\bibfnamefont {D.}~\bibnamefont
  {Mishra}}, \bibinfo {author} {\bibfnamefont {S.}~\bibnamefont {Singh}},
  \bibinfo {author} {\bibfnamefont {B.}~\bibnamefont {Kumar}},\ and\ \bibinfo
  {author} {\bibfnamefont {P.}~\bibnamefont {Jha}},\ }\href@noop {} {\bibfield
  {journal} {\bibinfo  {journal} {Physics of Plasmas}\ }\textbf {\bibinfo
  {volume} {32}} (\bibinfo {year} {2025})}\BibitemShut {NoStop}%
\bibitem [{\citenamefont {Yun}\ \emph {et~al.}(2025)\citenamefont {Yun},
  \citenamefont {Bae}, \citenamefont {Mirzaie},\ and\ \citenamefont
  {Kim}}]{yun2025laser}%
  \BibitemOpen
  \bibfield  {author} {\bibinfo {author} {\bibfnamefont {H.}~\bibnamefont
  {Yun}}, \bibinfo {author} {\bibfnamefont {L.~J.}\ \bibnamefont {Bae}},
  \bibinfo {author} {\bibfnamefont {M.}~\bibnamefont {Mirzaie}},\ and\ \bibinfo
  {author} {\bibfnamefont {H.~T.}\ \bibnamefont {Kim}},\ }\href@noop {}
  {\bibfield  {journal} {\bibinfo  {journal} {Reviews of Modern Plasma
  Physics}\ }\textbf {\bibinfo {volume} {9}},\ \bibinfo {pages} {13} (\bibinfo
  {year} {2025})}\BibitemShut {NoStop}%
\bibitem [{\citenamefont {Chamoli}\ \emph {et~al.}(2025)\citenamefont
  {Chamoli}, \citenamefont {Gupta}, \citenamefont {Singh},\ and\ \citenamefont
  {Kumar}}]{chamoli2025optical}%
  \BibitemOpen
  \bibfield  {author} {\bibinfo {author} {\bibfnamefont {A.}~\bibnamefont
  {Chamoli}}, \bibinfo {author} {\bibfnamefont {D.~N.}\ \bibnamefont {Gupta}},
  \bibinfo {author} {\bibfnamefont {A.~R.}\ \bibnamefont {Singh}},\ and\
  \bibinfo {author} {\bibfnamefont {V.}~\bibnamefont {Kumar}},\ }\href@noop {}
  {\bibfield  {journal} {\bibinfo  {journal} {Journal of Optics}\ ,\ \bibinfo
  {pages} {1}} (\bibinfo {year} {2025})}\BibitemShut {NoStop}%
\bibitem [{\citenamefont {Srivastav}\ and\ \citenamefont
  {Panwar}(2023{\natexlab{a}})}]{SrivastavNLMSH23}%
  \BibitemOpen
  \bibfield  {author} {\bibinfo {author} {\bibfnamefont {R.~K.}\ \bibnamefont
  {Srivastav}}\ and\ \bibinfo {author} {\bibfnamefont {A.}~\bibnamefont
  {Panwar}},\ }\href@noop {} {\bibfield  {journal} {\bibinfo  {journal}
  {Optical and Quantum Electronics}\ }\textbf {\bibinfo {volume} {55}},\
  \bibinfo {pages} {111} (\bibinfo {year} {2023}{\natexlab{a}})}\BibitemShut
  {NoStop}%
\bibitem [{\citenamefont {Zhang}\ \emph {et~al.}(2024)\citenamefont {Zhang},
  \citenamefont {Dong}, \citenamefont {Li}, \citenamefont {Cao}, \citenamefont
  {Yang}, \citenamefont {Yu}, \citenamefont {Yang}, \citenamefont {Wang},\ and\
  \citenamefont {Gong}}]{zhang2024plane}%
  \BibitemOpen
  \bibfield  {author} {\bibinfo {author} {\bibfnamefont {P.}~\bibnamefont
  {Zhang}}, \bibinfo {author} {\bibfnamefont {Y.}~\bibnamefont {Dong}},
  \bibinfo {author} {\bibfnamefont {X.}~\bibnamefont {Li}}, \bibinfo {author}
  {\bibfnamefont {X.}~\bibnamefont {Cao}}, \bibinfo {author} {\bibfnamefont
  {Y.}~\bibnamefont {Yang}}, \bibinfo {author} {\bibfnamefont {G.}~\bibnamefont
  {Yu}}, \bibinfo {author} {\bibfnamefont {S.}~\bibnamefont {Yang}}, \bibinfo
  {author} {\bibfnamefont {S.}~\bibnamefont {Wang}},\ and\ \bibinfo {author}
  {\bibfnamefont {Y.}~\bibnamefont {Gong}},\ }\href@noop {} {\bibfield
  {journal} {\bibinfo  {journal} {Micromachines}\ }\textbf {\bibinfo {volume}
  {15}},\ \bibinfo {pages} {723} (\bibinfo {year} {2024})}\BibitemShut
  {NoStop}%
\bibitem [{\citenamefont {Liu}, \citenamefont {Qian},\ and\ \citenamefont
  {Chong}(2015)}]{liu2015directional}%
  \BibitemOpen
  \bibfield  {author} {\bibinfo {author} {\bibfnamefont {F.}~\bibnamefont
  {Liu}}, \bibinfo {author} {\bibfnamefont {C.}~\bibnamefont {Qian}},\ and\
  \bibinfo {author} {\bibfnamefont {Y.~D.}\ \bibnamefont {Chong}},\ }\href@noop
  {} {\bibfield  {journal} {\bibinfo  {journal} {Optics express}\ }\textbf
  {\bibinfo {volume} {23}},\ \bibinfo {pages} {2383} (\bibinfo {year}
  {2015})}\BibitemShut {NoStop}%
\bibitem [{\citenamefont {Srivastav}\ and\ \citenamefont
  {Panwar}(2023{\natexlab{b}})}]{srivastav2023excitation}%
  \BibitemOpen
  \bibfield  {author} {\bibinfo {author} {\bibfnamefont {R.~K.}\ \bibnamefont
  {Srivastav}}\ and\ \bibinfo {author} {\bibfnamefont {A.}~\bibnamefont
  {Panwar}},\ }\href@noop {} {\bibfield  {journal} {\bibinfo  {journal}
  {Optical and Quantum Electronics}\ }\textbf {\bibinfo {volume} {55}},\
  \bibinfo {pages} {111} (\bibinfo {year} {2023}{\natexlab{b}})}\BibitemShut
  {NoStop}%
\bibitem [{\citenamefont {Korani}\ \emph {et~al.}(2024)\citenamefont {Korani},
  \citenamefont {Mohammadi}, \citenamefont {Hocini},\ and\ \citenamefont
  {Danaie}}]{korani2024tunable}%
  \BibitemOpen
  \bibfield  {author} {\bibinfo {author} {\bibfnamefont {N.}~\bibnamefont
  {Korani}}, \bibinfo {author} {\bibfnamefont {S.}~\bibnamefont {Mohammadi}},
  \bibinfo {author} {\bibfnamefont {A.}~\bibnamefont {Hocini}},\ and\ \bibinfo
  {author} {\bibfnamefont {M.}~\bibnamefont {Danaie}},\ }\href@noop {}
  {\bibfield  {journal} {\bibinfo  {journal} {Diamond and Related Materials}\
  }\textbf {\bibinfo {volume} {149}},\ \bibinfo {pages} {111554} (\bibinfo
  {year} {2024})}\BibitemShut {NoStop}%
\bibitem [{\citenamefont {Vatoor}, \citenamefont {Tabatabaee},\ and\
  \citenamefont {Shabani}(2025)}]{vatoor2025tunable}%
  \BibitemOpen
  \bibfield  {author} {\bibinfo {author} {\bibfnamefont {M.}~\bibnamefont
  {Vatoor}}, \bibinfo {author} {\bibfnamefont {S.~S.}\ \bibnamefont
  {Tabatabaee}},\ and\ \bibinfo {author} {\bibfnamefont {P.}~\bibnamefont
  {Shabani}},\ }\href@noop {} {\bibfield  {journal} {\bibinfo  {journal}
  {Sensing and Bio-Sensing Research}\ }\textbf {\bibinfo {volume} {47}},\
  \bibinfo {pages} {100776} (\bibinfo {year} {2025})}\BibitemShut {NoStop}%
\bibitem [{\citenamefont {Ukirade}(2025)}]{ukirade2025review}%
  \BibitemOpen
  \bibfield  {author} {\bibinfo {author} {\bibfnamefont {N.~A.}\ \bibnamefont
  {Ukirade}},\ }\href@noop {} {\bibfield  {journal} {\bibinfo  {journal} {Next
  Materials}\ }\textbf {\bibinfo {volume} {6}},\ \bibinfo {pages} {100479}
  (\bibinfo {year} {2025})}\BibitemShut {NoStop}%
\bibitem [{\citenamefont {Hasan}\ \emph {et~al.}(2024)\citenamefont {Hasan},
  \citenamefont {Khan}, \citenamefont {Shahzadi}, \citenamefont {Bagheri},\
  and\ \citenamefont {Ban}}]{hasan2024terahertz}%
  \BibitemOpen
  \bibfield  {author} {\bibinfo {author} {\bibfnamefont {M.~S.}\ \bibnamefont
  {Hasan}}, \bibinfo {author} {\bibfnamefont {A.~A.}\ \bibnamefont {Khan}},
  \bibinfo {author} {\bibfnamefont {S.}~\bibnamefont {Shahzadi}}, \bibinfo
  {author} {\bibfnamefont {M.~H.}\ \bibnamefont {Bagheri}},\ and\ \bibinfo
  {author} {\bibfnamefont {D.}~\bibnamefont {Ban}},\ }\href@noop {} {\bibfield
  {journal} {\bibinfo  {journal} {Advanced Functional Materials}\ }\textbf
  {\bibinfo {volume} {34}},\ \bibinfo {pages} {2400313} (\bibinfo {year}
  {2024})}\BibitemShut {NoStop}%
\bibitem [{\citenamefont {Kumar}\ \emph {et~al.}(2022)\citenamefont {Kumar},
  \citenamefont {Vij}, \citenamefont {Kant},\ and\ \citenamefont
  {Thakur}}]{kumar2022combined}%
  \BibitemOpen
  \bibfield  {author} {\bibinfo {author} {\bibfnamefont {S.}~\bibnamefont
  {Kumar}}, \bibinfo {author} {\bibfnamefont {S.}~\bibnamefont {Vij}}, \bibinfo
  {author} {\bibfnamefont {N.}~\bibnamefont {Kant}},\ and\ \bibinfo {author}
  {\bibfnamefont {V.}~\bibnamefont {Thakur}},\ }\href@noop {} {\bibfield
  {journal} {\bibinfo  {journal} {Journal of Astrophysics and Astronomy}\
  }\textbf {\bibinfo {volume} {43}},\ \bibinfo {pages} {30} (\bibinfo {year}
  {2022})}\BibitemShut {NoStop}%
\bibitem [{\citenamefont {Ghayemmoniri}, \citenamefont {Siahmazgi},\ and\
  \citenamefont {Jafari}(2023)}]{ghayemmoniri2023terahertz}%
  \BibitemOpen
  \bibfield  {author} {\bibinfo {author} {\bibfnamefont {Z.}~\bibnamefont
  {Ghayemmoniri}}, \bibinfo {author} {\bibfnamefont {R.~N.}\ \bibnamefont
  {Siahmazgi}},\ and\ \bibinfo {author} {\bibfnamefont {S.}~\bibnamefont
  {Jafari}},\ }\href@noop {} {\bibfield  {journal} {\bibinfo  {journal} {The
  European Physical Journal D}\ }\textbf {\bibinfo {volume} {77}},\ \bibinfo
  {pages} {48} (\bibinfo {year} {2023})}\BibitemShut {NoStop}%
\bibitem [{\citenamefont {Sepehri~Javan}\ and\ \citenamefont
  {Rouhi~Erdi}(2017)}]{sepehri2017theoretical}%
  \BibitemOpen
  \bibfield  {author} {\bibinfo {author} {\bibfnamefont {N.}~\bibnamefont
  {Sepehri~Javan}}\ and\ \bibinfo {author} {\bibfnamefont {F.}~\bibnamefont
  {Rouhi~Erdi}},\ }\href@noop {} {\bibfield  {journal} {\bibinfo  {journal}
  {Journal of Applied Physics}\ }\textbf {\bibinfo {volume} {122}} (\bibinfo
  {year} {2017})}\BibitemShut {NoStop}%
\bibitem [{\citenamefont {Rahmanpour~Kolur}\ and\ \citenamefont
  {Esmaeilzadeh}(2025)}]{rahmanpour2025terahertz}%
  \BibitemOpen
  \bibfield  {author} {\bibinfo {author} {\bibfnamefont {E.}~\bibnamefont
  {Rahmanpour~Kolur}}\ and\ \bibinfo {author} {\bibfnamefont {M.}~\bibnamefont
  {Esmaeilzadeh}},\ }\href@noop {} {\bibfield  {journal} {\bibinfo  {journal}
  {Physics of Plasmas}\ }\textbf {\bibinfo {volume} {32}} (\bibinfo {year}
  {2025})}\BibitemShut {NoStop}%
\bibitem [{\citenamefont {Kumar}\ \emph {et~al.}(2024)\citenamefont {Kumar},
  \citenamefont {Gopal}, \citenamefont {Singh},\ and\ \citenamefont
  {Singh}}]{10456555}%
  \BibitemOpen
  \bibfield  {author} {\bibinfo {author} {\bibfnamefont {R.}~\bibnamefont
  {Kumar}}, \bibinfo {author} {\bibfnamefont {K.}~\bibnamefont {Gopal}},
  \bibinfo {author} {\bibfnamefont {D.}~\bibnamefont {Singh}},\ and\ \bibinfo
  {author} {\bibfnamefont {S.}~\bibnamefont {Singh}},\ }\href
  {https://doi.org/10.1109/TPS.2024.3365824} {\bibfield  {journal} {\bibinfo
  {journal} {IEEE Transactions on Plasma Science}\ }\textbf {\bibinfo {volume}
  {52}},\ \bibinfo {pages} {1053} (\bibinfo {year} {2024})}\BibitemShut
  {NoStop}%
\bibitem [{\citenamefont {Singh}\ \emph {et~al.}(2024)\citenamefont {Singh},
  \citenamefont {Gopal}, \citenamefont {Gupta}, \citenamefont {Kundu},\ and\
  \citenamefont {Varshney}}]{singh2024laser}%
  \BibitemOpen
  \bibfield  {author} {\bibinfo {author} {\bibfnamefont {A.}~\bibnamefont
  {Singh}}, \bibinfo {author} {\bibfnamefont {K.}~\bibnamefont {Gopal}},
  \bibinfo {author} {\bibfnamefont {D.}~\bibnamefont {Gupta}}, \bibinfo
  {author} {\bibfnamefont {M.}~\bibnamefont {Kundu}},\ and\ \bibinfo {author}
  {\bibfnamefont {P.}~\bibnamefont {Varshney}},\ }\href@noop {} {\bibfield
  {journal} {\bibinfo  {journal} {Indian Journal of Physics}\ }\textbf
  {\bibinfo {volume} {98}},\ \bibinfo {pages} {383} (\bibinfo {year}
  {2024})}\BibitemShut {NoStop}%
\bibitem [{\citenamefont {Srivastav}\ and\ \citenamefont
  {Panwar}(2022{\natexlab{a}})}]{srivastav2022excitation}%
  \BibitemOpen
  \bibfield  {author} {\bibinfo {author} {\bibfnamefont {R.~K.}\ \bibnamefont
  {Srivastav}}\ and\ \bibinfo {author} {\bibfnamefont {A.}~\bibnamefont
  {Panwar}},\ }\href@noop {} {\bibfield  {journal} {\bibinfo  {journal}
  {Optik}\ }\textbf {\bibinfo {volume} {264}},\ \bibinfo {pages} {169363}
  (\bibinfo {year} {2022}{\natexlab{a}})}\BibitemShut {NoStop}%
\bibitem [{\citenamefont {Srivastav}\ and\ \citenamefont
  {Kundu}(2026)}]{srivastav2025terahertz}%
  \BibitemOpen
  \bibfield  {author} {\bibinfo {author} {\bibfnamefont {R.~K.}\ \bibnamefont
  {Srivastav}}\ and\ \bibinfo {author} {\bibfnamefont {M.}~\bibnamefont
  {Kundu}},\ }\href {https://doi.org/10.1103/lfjf-g5n5} {\bibfield  {journal}
  {\bibinfo  {journal} {Phys. Rev. E}\ }\textbf {\bibinfo {volume} {113}},\
  \bibinfo {pages} {015208} (\bibinfo {year} {2026})}\BibitemShut {NoStop}%
\bibitem [{\citenamefont {Singh}\ and\ \citenamefont
  {Tripathi}(2007)}]{singh2007surface}%
  \BibitemOpen
  \bibfield  {author} {\bibinfo {author} {\bibfnamefont {D.}~\bibnamefont
  {Singh}}\ and\ \bibinfo {author} {\bibfnamefont {V.}~\bibnamefont
  {Tripathi}},\ }\href@noop {} {\bibfield  {journal} {\bibinfo  {journal}
  {Journal of Applied Physics}\ }\textbf {\bibinfo {volume} {102}} (\bibinfo
  {year} {2007})}\BibitemShut {NoStop}%
\bibitem [{\citenamefont {Bhasin}\ and\ \citenamefont
  {Tripathi}(2010)}]{bhasin2010resonant}%
  \BibitemOpen
  \bibfield  {author} {\bibinfo {author} {\bibfnamefont {L.}~\bibnamefont
  {Bhasin}}\ and\ \bibinfo {author} {\bibfnamefont {V.~K.}\ \bibnamefont
  {Tripathi}},\ }\href@noop {} {\bibfield  {journal} {\bibinfo  {journal} {IEEE
  Journal of Quantum Electronics}\ }\textbf {\bibinfo {volume} {46}},\ \bibinfo
  {pages} {965} (\bibinfo {year} {2010})}\BibitemShut {NoStop}%
\bibitem [{\citenamefont {Yarahmadi}, \citenamefont {Moravvej-Farshi},\ and\
  \citenamefont {Yousefi}(2015)}]{yarahmadi2015subwavelength}%
  \BibitemOpen
  \bibfield  {author} {\bibinfo {author} {\bibfnamefont {M.}~\bibnamefont
  {Yarahmadi}}, \bibinfo {author} {\bibfnamefont {M.~K.}\ \bibnamefont
  {Moravvej-Farshi}},\ and\ \bibinfo {author} {\bibfnamefont {L.}~\bibnamefont
  {Yousefi}},\ }\href@noop {} {\bibfield  {journal} {\bibinfo  {journal} {IEEE
  Transactions on Terahertz Science and Technology}\ }\textbf {\bibinfo
  {volume} {5}},\ \bibinfo {pages} {725} (\bibinfo {year} {2015})}\BibitemShut
  {NoStop}%
\bibitem [{\citenamefont {Liu}\ \emph {et~al.}(2011)\citenamefont {Liu},
  \citenamefont {Yin}, \citenamefont {Ulin-Avila}, \citenamefont {Geng},
  \citenamefont {Zentgraf}, \citenamefont {Ju}, \citenamefont {Wang},\ and\
  \citenamefont {Zhang}}]{liu2011graphene}%
  \BibitemOpen
  \bibfield  {author} {\bibinfo {author} {\bibfnamefont {M.}~\bibnamefont
  {Liu}}, \bibinfo {author} {\bibfnamefont {X.}~\bibnamefont {Yin}}, \bibinfo
  {author} {\bibfnamefont {E.}~\bibnamefont {Ulin-Avila}}, \bibinfo {author}
  {\bibfnamefont {B.}~\bibnamefont {Geng}}, \bibinfo {author} {\bibfnamefont
  {T.}~\bibnamefont {Zentgraf}}, \bibinfo {author} {\bibfnamefont
  {L.}~\bibnamefont {Ju}}, \bibinfo {author} {\bibfnamefont {F.}~\bibnamefont
  {Wang}},\ and\ \bibinfo {author} {\bibfnamefont {X.}~\bibnamefont {Zhang}},\
  }\href@noop {} {\bibfield  {journal} {\bibinfo  {journal} {Nature}\ }\textbf
  {\bibinfo {volume} {474}},\ \bibinfo {pages} {64} (\bibinfo {year}
  {2011})}\BibitemShut {NoStop}%
\bibitem [{\citenamefont {Brion}\ \emph {et~al.}(1972)\citenamefont {Brion},
  \citenamefont {Wallis}, \citenamefont {Hartstein},\ and\ \citenamefont
  {Burstein}}]{brion1972theory}%
  \BibitemOpen
  \bibfield  {author} {\bibinfo {author} {\bibfnamefont {J.}~\bibnamefont
  {Brion}}, \bibinfo {author} {\bibfnamefont {R.}~\bibnamefont {Wallis}},
  \bibinfo {author} {\bibfnamefont {A.}~\bibnamefont {Hartstein}},\ and\
  \bibinfo {author} {\bibfnamefont {E.}~\bibnamefont {Burstein}},\ }\href@noop
  {} {\bibfield  {journal} {\bibinfo  {journal} {Physical Review Letters}\
  }\textbf {\bibinfo {volume} {28}},\ \bibinfo {pages} {1455} (\bibinfo {year}
  {1972})}\BibitemShut {NoStop}%
\bibitem [{\citenamefont {Jing}\ \emph {et~al.}(2022)\citenamefont {Jing},
  \citenamefont {Wei}, \citenamefont {Duan}, \citenamefont {Hao}, \citenamefont
  {Zhao}, \citenamefont {Qu}, \citenamefont {Wang},\ and\ \citenamefont
  {Zhang}}]{jing2022thermally}%
  \BibitemOpen
  \bibfield  {author} {\bibinfo {author} {\bibfnamefont {H.}~\bibnamefont
  {Jing}}, \bibinfo {author} {\bibfnamefont {Y.}~\bibnamefont {Wei}}, \bibinfo
  {author} {\bibfnamefont {J.}~\bibnamefont {Duan}}, \bibinfo {author}
  {\bibfnamefont {J.}~\bibnamefont {Hao}}, \bibinfo {author} {\bibfnamefont
  {W.}~\bibnamefont {Zhao}}, \bibinfo {author} {\bibfnamefont {Z.}~\bibnamefont
  {Qu}}, \bibinfo {author} {\bibfnamefont {J.}~\bibnamefont {Wang}},\ and\
  \bibinfo {author} {\bibfnamefont {B.}~\bibnamefont {Zhang}},\ }\href@noop {}
  {\bibfield  {journal} {\bibinfo  {journal} {Optical Materials}\ }\textbf
  {\bibinfo {volume} {129}},\ \bibinfo {pages} {112311} (\bibinfo {year}
  {2022})}\BibitemShut {NoStop}%
\bibitem [{\citenamefont {Gao}\ \emph {et~al.}(2023)\citenamefont {Gao},
  \citenamefont {Sun}, \citenamefont {Li}, \citenamefont {Su}, \citenamefont
  {Sun}, \citenamefont {Xia}, \citenamefont {Zhang}, \citenamefont {Dong},\
  and\ \citenamefont {Yun}}]{gao2023multifunctional}%
  \BibitemOpen
  \bibfield  {author} {\bibinfo {author} {\bibfnamefont {P.}~\bibnamefont
  {Gao}}, \bibinfo {author} {\bibfnamefont {J.}~\bibnamefont {Sun}}, \bibinfo
  {author} {\bibfnamefont {W.}~\bibnamefont {Li}}, \bibinfo {author}
  {\bibfnamefont {C.}~\bibnamefont {Su}}, \bibinfo {author} {\bibfnamefont
  {Z.}~\bibnamefont {Sun}}, \bibinfo {author} {\bibfnamefont {F.}~\bibnamefont
  {Xia}}, \bibinfo {author} {\bibfnamefont {K.}~\bibnamefont {Zhang}}, \bibinfo
  {author} {\bibfnamefont {L.}~\bibnamefont {Dong}},\ and\ \bibinfo {author}
  {\bibfnamefont {M.}~\bibnamefont {Yun}},\ }\href@noop {} {\bibfield
  {journal} {\bibinfo  {journal} {Results in Physics}\ }\textbf {\bibinfo
  {volume} {52}},\ \bibinfo {pages} {106797} (\bibinfo {year}
  {2023})}\BibitemShut {NoStop}%
\bibitem [{\citenamefont {Chen}(1984)}]{chen1984introduction}%
  \BibitemOpen
  \bibfield  {author} {\bibinfo {author} {\bibfnamefont {F.~F.}\ \bibnamefont
  {Chen}},\ }\href@noop {} {\emph {\bibinfo {title} {Introduction to plasma
  physics and controlled fusion}}},\ Vol.~\bibinfo {volume} {1}\ (\bibinfo
  {publisher} {Springer},\ \bibinfo {year} {1984})\BibitemShut {NoStop}%
\bibitem [{\citenamefont {Liu}, \citenamefont {Tripathi},\ and\ \citenamefont
  {Eliasson}(2019)}]{liu2019high}%
  \BibitemOpen
  \bibfield  {author} {\bibinfo {author} {\bibfnamefont {C.}~\bibnamefont
  {Liu}}, \bibinfo {author} {\bibfnamefont {V.~K.}\ \bibnamefont {Tripathi}},\
  and\ \bibinfo {author} {\bibfnamefont {B.}~\bibnamefont {Eliasson}},\
  }\href@noop {} {\emph {\bibinfo {title} {High-power laser-plasma
  interaction}}}\ (\bibinfo  {publisher} {Cambridge university press},\
  \bibinfo {year} {2019})\BibitemShut {NoStop}%
\bibitem [{\citenamefont {Srivastav}\ and\ \citenamefont
  {Panwar}(2022{\natexlab{b}})}]{Srivastav2022}%
  \BibitemOpen
  \bibfield  {author} {\bibinfo {author} {\bibfnamefont {R.~K.}\ \bibnamefont
  {Srivastav}}\ and\ \bibinfo {author} {\bibfnamefont {A.}~\bibnamefont
  {Panwar}},\ }\href
  {https://doi.org/https://doi.org/10.1016/j.ijleo.2022.169363} {\bibfield
  {journal} {\bibinfo  {journal} {Optik}\ }\textbf {\bibinfo {volume} {264}},\
  \bibinfo {pages} {169363} (\bibinfo {year} {2022}{\natexlab{b}})}\BibitemShut
  {NoStop}%
\bibitem [{\citenamefont {Liu}\ and\ \citenamefont
  {Tripathi}(2017)}]{10.1117/1.JNP.11.036015}%
  \BibitemOpen
  \bibfield  {author} {\bibinfo {author} {\bibfnamefont {C.~S.}\ \bibnamefont
  {Liu}}\ and\ \bibinfo {author} {\bibfnamefont {V.~K.}\ \bibnamefont
  {Tripathi}},\ }\href@noop {} {\bibfield  {journal} {\bibinfo  {journal}
  {Journal of Nanophotonics}\ }\textbf {\bibinfo {volume} {11}},\ \bibinfo
  {pages} {036015} (\bibinfo {year} {2017})}\BibitemShut {NoStop}%
\bibitem [{\citenamefont {Verma}, \citenamefont {Govindan},\ and\ \citenamefont
  {Kumar}(2021)}]{verma2021terahertz}%
  \BibitemOpen
  \bibfield  {author} {\bibinfo {author} {\bibfnamefont {N.}~\bibnamefont
  {Verma}}, \bibinfo {author} {\bibfnamefont {A.}~\bibnamefont {Govindan}},\
  and\ \bibinfo {author} {\bibfnamefont {P.}~\bibnamefont {Kumar}},\
  }\href@noop {} {\bibfield  {journal} {\bibinfo  {journal} {Plasmonics}\
  }\textbf {\bibinfo {volume} {16}},\ \bibinfo {pages} {711} (\bibinfo {year}
  {2021})}\BibitemShut {NoStop}%
\bibitem [{\citenamefont {Son}, \citenamefont {Oh},\ and\ \citenamefont
  {Cheon}(2019)}]{son2019potential}%
  \BibitemOpen
  \bibfield  {author} {\bibinfo {author} {\bibfnamefont {J.-H.}\ \bibnamefont
  {Son}}, \bibinfo {author} {\bibfnamefont {S.~J.}\ \bibnamefont {Oh}},\ and\
  \bibinfo {author} {\bibfnamefont {H.}~\bibnamefont {Cheon}},\ }\href@noop {}
  {\bibfield  {journal} {\bibinfo  {journal} {Journal of Applied Physics}\
  }\textbf {\bibinfo {volume} {125}} (\bibinfo {year} {2019})}\BibitemShut
  {NoStop}%
\bibitem [{\citenamefont {Nikitkina}\ \emph {et~al.}(2021)\citenamefont
  {Nikitkina}, \citenamefont {Bikmulina}, \citenamefont {Gafarova},
  \citenamefont {Kosheleva}, \citenamefont {Efremov}, \citenamefont {Bezrukov},
  \citenamefont {Butnaru}, \citenamefont {Dolganova}, \citenamefont
  {Chernomyrdin}, \citenamefont {Cherkasova}, \citenamefont {Gavdush},\ and\
  \citenamefont {Timashev}}]{Angelina2021}%
  \BibitemOpen
  \bibfield  {author} {\bibinfo {author} {\bibfnamefont {A.~I.}\ \bibnamefont
  {Nikitkina}}, \bibinfo {author} {\bibfnamefont {P.}~\bibnamefont
  {Bikmulina}}, \bibinfo {author} {\bibfnamefont {E.~R.}\ \bibnamefont
  {Gafarova}}, \bibinfo {author} {\bibfnamefont {N.~V.}\ \bibnamefont
  {Kosheleva}}, \bibinfo {author} {\bibfnamefont {Y.~M.}\ \bibnamefont
  {Efremov}}, \bibinfo {author} {\bibfnamefont {E.~A.}\ \bibnamefont
  {Bezrukov}}, \bibinfo {author} {\bibfnamefont {D.~V.}\ \bibnamefont
  {Butnaru}}, \bibinfo {author} {\bibfnamefont {I.~N.}\ \bibnamefont
  {Dolganova}}, \bibinfo {author} {\bibfnamefont {N.~V.}\ \bibnamefont
  {Chernomyrdin}}, \bibinfo {author} {\bibfnamefont {O.~P.}\ \bibnamefont
  {Cherkasova}}, \bibinfo {author} {\bibfnamefont {A.~A.}\ \bibnamefont
  {Gavdush}},\ and\ \bibinfo {author} {\bibfnamefont {P.~S.}\ \bibnamefont
  {Timashev}},\ }\href@noop {} {\bibfield  {journal} {\bibinfo  {journal}
  {Journal of Biomedical Optics}\ }\textbf {\bibinfo {volume} {26}},\ \bibinfo
  {pages} {043005} (\bibinfo {year} {2021})}\BibitemShut {NoStop}%
\bibitem [{\citenamefont {Salameh}\ and\ \citenamefont
  {El~Tarhuni}(2022)}]{salameh20225g}%
  \BibitemOpen
  \bibfield  {author} {\bibinfo {author} {\bibfnamefont {A.~I.}\ \bibnamefont
  {Salameh}}\ and\ \bibinfo {author} {\bibfnamefont {M.}~\bibnamefont
  {El~Tarhuni}},\ }\href@noop {} {\bibfield  {journal} {\bibinfo  {journal}
  {Future Internet}\ }\textbf {\bibinfo {volume} {14}},\ \bibinfo {pages} {117}
  (\bibinfo {year} {2022})}\BibitemShut {NoStop}%
\bibitem [{\citenamefont {Kaur}\ \emph {et~al.}(2024)\citenamefont {Kaur},
  \citenamefont {Islam}, \citenamefont {Agarwal}, \citenamefont {Kaur},\ and\
  \citenamefont {Kumar}}]{kaur2024highly}%
  \BibitemOpen
  \bibfield  {author} {\bibinfo {author} {\bibfnamefont {R.}~\bibnamefont
  {Kaur}}, \bibinfo {author} {\bibfnamefont {M.}~\bibnamefont {Islam}},
  \bibinfo {author} {\bibfnamefont {P.}~\bibnamefont {Agarwal}}, \bibinfo
  {author} {\bibfnamefont {S.}~\bibnamefont {Kaur}},\ and\ \bibinfo {author}
  {\bibfnamefont {G.}~\bibnamefont {Kumar}},\ }\href@noop {} {\bibfield
  {journal} {\bibinfo  {journal} {Journal of Optics}\ }\textbf {\bibinfo
  {volume} {53}},\ \bibinfo {pages} {2955} (\bibinfo {year}
  {2024})}\BibitemShut {NoStop}%
\end{thebibliography}%

\end{document}